\documentclass[useAMS,usenatbib]{mn2e}
\usepackage{subfigure}
\usepackage{xspace}
\usepackage{graphicx, amssymb}
\usepackage[fleqn]{amsmath}

\title[First stars under X-ray feedback]{The First Stars: formation under X-ray feedback}

\author[J.~A. Hummel et al.]
{Jacob~A.~Hummel$^1$\thanks{E-mail: jhummel@astro.as.utexas.edu},
  Athena~Stacy$^2$, Myoungwon~Jeon$^1$, Anthony~Oliveri$^1$
  \newauthor and Volker~Bromm$^1$\\
$^1$Department of Astronomy, The University of Texas at Austin, TX 78712, USA\\
$^2$University of California, Berkeley, CA 94720, USA
 }

\newcommand{\kelvin}{\ensuremath{\,{\rm K}}\xspace}
\newcommand{\s}{\ensuremath{\,{\rm s}}\xspace}

\newcommand{\cm}{\ensuremath{\,{\rm cm}}\xspace}
\newcommand{\cc}{\ensuremath{\,{\rm cm}^{-3}}\xspace}
\newcommand{\msun}{\ensuremath{\,{\rm M}_{\odot}}\xspace}
\newcommand{\rsun}{\ensuremath{\,{\rm R}_{\odot}}\xspace}

\newcommand{\kms}{\ensuremath{\,\mathrm{km}\,\mathrm{s}^{-1}\xspace}}
\newcommand{\pc}{\ensuremath{\,{\rm pc}}\xspace}
\newcommand{\au}{\ensuremath{\,{\rm AU}}\xspace}
\newcommand{\Mpc}{\ensuremath{\,{\rm Mpc}}\xspace}

\newcommand{\yr}{\ensuremath{\,{\rm yr}}\xspace}

\newcommand{\ev}{\ensuremath{\,{\rm eV}}\xspace}
\newcommand{\kev}{\ensuremath{\,{\rm keV}}\xspace}
\newcommand{\erg}{\ensuremath{\,{\rm erg}}\xspace}
\newcommand{\sr}{\ensuremath{\,{\rm sr}}\xspace}

\newcommand{\tff}{\ensuremath{{t}_{\mathrm{ff}}}\xspace}
\newcommand{\tH}{\ensuremath{{t}_{\textsc{h}}}\xspace}

\newcommand{\rvir}{\ensuremath{{R}_{\mathrm{vir}}}\xspace}
\newcommand{\mvir}{\ensuremath{{M}_{\mathrm{vir}}}\xspace}

\newcommand{\htwo}{\ensuremath{\mathrm{H}_2}\xspace}
\newcommand{\hd}{\ensuremath{\mathrm{HD}}\xspace}
\newcommand{\deut}{\ensuremath{\mathrm{D}}\xspace}
\newcommand{\h}{\ensuremath{\mathrm{H}}\xspace}
\newcommand{\hplus}{\ensuremath{\mathrm{H}^+}\xspace}
\newcommand{\hminus}{\ensuremath{\mathrm{H}^-}\xspace}
\newcommand{\he}{\ensuremath{\mathrm{He}}\xspace}
\newcommand{\heplus}{\ensuremath{\mathrm{He}^+}\xspace}

\newcommand{\HI}{\ensuremath{\mathrm{H\,\textsc{i}}}\xspace}
\newcommand{\HeI}{\ensuremath{\mathrm{He\,\textsc{i}}}\xspace}
\newcommand{\HeII}{\ensuremath{\mathrm{He\,\textsc{ii}}}\xspace}

\newcommand{\egrav}{\ensuremath{{E}_{\mathrm{grav}}}\xspace}
\newcommand{\etherm}{\ensuremath{{E}_{\mathrm{th}}}\xspace}

\newcommand{\uxr}{\ensuremath{{u}_{\textsc{xr}}}\xspace}
\newcommand{\uxrz}{\ensuremath{{u}_{\textsc{xr}}(z)}\xspace}
\newcommand{\gxr}{\ensuremath{\Gamma_{\textsc{xr}}}\xspace}
\newcommand{\gcrit}{\ensuremath{\Gamma_{\textsc{xr}, \mathrm{crit}}}\xspace}

\newcommand{\Lxr}{\ensuremath{L_{\textsc{xr}}}\xspace}
\newcommand{\jxr}{\ensuremath{J_{\textsc{xr}}}\xspace}
\newcommand{\jxrz}{\ensuremath{J_{\textsc{xr}}(z)}\xspace}
\newcommand{\jxrvz}{\ensuremath{J_{\nu, \textsc{xr}}(z)}\xspace}
\newcommand{\jcrit}{\ensuremath{J_{\textsc{xr}, \mathrm{crit}}}\xspace}

\newcommand{\sfrz}{\ensuremath{\Psi_{*}(z)}\xspace}
\newcommand{\xrb}{\ensuremath{\textsc{hmxb}}\xspace}

\newcommand{\about}{\ensuremath{\sim}}
\newcommand{\RefTab}[1]{\mbox{Table~\ref{#1}}}                     
\newcommand{\RefFig}[1]{\mbox{Figure~\ref{#1}}}

\newcommand{\RefSec}[1]{\mbox{Section~\ref{#1}}}                        

\begin{document}

\label{firstpage}

\maketitle
\topmargin-1cm

\begin{abstract}
We investigate the impact of a cosmic X-ray background (CXB) on Population III stars forming in a minihalo at $z\simeq25$.  Using the smoothed particle hydrodynamics code \textsc{gadget-2}, we attain sufficient numerical resolution to follow gas collapsing into the centre of the minihalo  from cosmological initial conditions up to densities of $10^{12}\cc$, at which point we form sink particles.  This allows us to study how the presence of a CXB affects the formation of \htwo and \hd in the gas prior to becoming fully molecular.   Using a suite of simulations for a range of possible CXB models, we follow each simulation for 5000\yr after the first sink particle forms.  The CXB provides two competing effects, with X-rays both heating the gas and increasing the free electron fraction, allowing more H$_2$ to form.  X-ray heating dominates below $n\sim1\cc$, while the additional H$_2$ cooling becomes more important above $n\sim10^2\cc$.  The gas becomes optically thick to X-rays as it exits the quasi-hydrostatic `loitering phase,' such that the primary impact of the CXB is to cool the gas at intermediate densities, resulting in an earlier onset of baryonic collapse into the dark matter halo. At the highest densities, self-shielding results in similar thermodynamic behaviour across a wide range of CXB strengths. Consequently, we find that star formation is relatively insensitive to the presence of a CXB; both the number and the characteristic mass of the stars formed remains quite similar even as the strength of the CXB varies by several orders of magnitude.
\end{abstract}


\section{Introduction}
\label{intro}
The formation of the first stars in the Universe marked a watershed moment in cosmic history.  It was during this as-yet unobserved epoch that our Universe began its transformation from the relatively simple initial conditions of the Big Bang to the complex tapestry of dark matter (DM), galaxies, stars and planets that we see today \citep{BarkanaLoeb2001, Miralda-Escude2003, Brommetal2009, Loeb2010}.  The radiation from these so-called Population III (Pop III) stars swept through the Universe, beginning the process of reionisation \citep{Kitayamaetal2004, Sokasianetal2004, WhalenAbelNorman2004, AlvarezBrommShapiro2006, JohnsonGreifBromm2007, Robertsonetal2010}, while the heavy elements forged in their cores and released in the violent supernova explosions marking their deaths began the process of chemical enrichment (\citealt{MadauFerraraRees2001, MoriFerraraMadau2002, BrommYoshidaHernquist2003, Hegeretal2003, UmedaNomoto2003, TornatoreFerraraSchneider2007, Greifetal2007, Greifetal2010, WiseAbel2008, Maioetal2011}; recently reviewed in \citealt{KarlssonBrommHawthorn2013}). These effects are strongly dependent on the characteristic mass of Pop III stars, which determines their total luminosity and ionising radiation output \citep{Schaerer2002}, and the details of their eventual demise \citep{Hegeretal2003, HegerWoosley2010, MaederMeynet2012}. As a result, developing a thorough knowledge of how environmental effects influence the properties of these stars is crucial to understanding their impact on the intergalactic medium (IGM) and subsequent stellar generations.

While the complexities of the various physical processes involved have so far prevented a definitive answer to this question, the basic properties of the first stars have been fairly well established, with the consensus that they formed in dark matter `minihaloes,' having on the order of $10^5 - 10^6 \msun$ at $z\gtrsim 20$ \citep{CouchmanRees1986, HaimanThoulLoeb1996, Tegmarketal1997}.  Pioneering numerical studies of the collapse of metal-free gas into these haloes, where molecular hydrogen was the only available coolant, suggested that Pop III stars were very massive---on the order of 100\msun---due to the lack of coolants more efficient than \htwo \citep[e.g.,][]{BrommCoppiLarson1999, BrommCoppiLarson2002, AbelBryanNorman2002, Yoshidaetal2003, BrommLarson2004, Yoshidaetal2006, O'SheaNorman2007}.  More recent simulations, benefiting from increased resolution, have found that significant fragmentation of the protostellar disc occurs during the star formation process, with the protostellar cores ranging in mass from \about0.1 to tens of solar masses \citep{StacyGreifBromm2010, Clarketal2011a, Clarketal2011b, Greifetal2011, Greifetal2012, StacyBromm2013, SusaHasegawaTominaga2014, Hiranoetal2014,Hiranoetal2015}, with a presumably flat initial mass function (IMF; \citealt{Dopckeetal2013}).

One intriguing outcome of these studies is that while the protostellar disc does indeed fragment, it only marginally satisfies the \citet{Gammie2001} criterion for disc instability \citep{Clarketal2011b, Greifetal2011, Greifetal2012}. While protostellar feedback is insufficient to stabilise the disc \citep{Smithetal2011, StacyGreifBromm2012}, it is possible that an external heating source could serve to stabilise the disc and prevent fragmentation. One promising source of such an external background is far-ultraviolet radiation in the Lyman-Werner (LW) bands (11.2-13.6 \ev).  While lacking sufficient energy to interact with atomic hydrogen, LW photons efficiently dissociate \htwo molecules, which serve as the primary coolant in primordial gas.  This diminishes the ability of the gas to cool, but studies have found that the critical LW flux required to suppress \htwo cooling is far above the expected mean value of such radiation \citep{Dijkstraetal2008}. Another possible heating source was recently explored by \citeauthor{Smithetal2012b}  (\citeyear{Smithetal2012b}; see also \citealt{Ripamontietal2009, Ripamontietal2010}) who investigated the ability of DM annihilation  to suppress fragmentation of the protostellar disc.  While such heating is unable to suppress star formation, it does serve to stabilise the disc, suppressing fragmentation within 1000 AU of the central protostar, at least for a while \citep{Stacyetal2012, Stacyetal2014}.

While LW radiation alone is unable to reliably suppress \htwo cooling in minihaloes, significant sources of LW photons, i.e., active star forming regions, contain large numbers of massive stars, and possibly mini-quasars as well \citep{KuhlenMadau2005, Jeonetal2012, Jeonetal2014a}.  Not only do massive stars end their lives as supernovae, leaving behind remnants that are significant sources of X-rays, a significant fraction of these stars are likely to be in tight binaries \citep[e.g.,][]{Clarketal2011b,Greifetal2012, Mirocha2014} and produce high-mass X-ray binaries (HMXBs).  As the cross-section of neutral hydrogen for X-rays is relatively small, such photons easily escape their host haloes, building up a cosmic X-ray background (CXB; \citealt{Oh2001, HaimanAbelRees2000, VenkatesanGirouxShull2001, GloverBrand2003, Cen2003, KuhlenMadau2005, Jeonetal2012, Jeonetal2014a}). This CXB serves to both heat and increase the ionisation fraction of gas in neighbouring minihaloes, which in turn can serve to increase the \htwo fraction of the gas by increasing the number of free electrons available to act as catalysts.

In this paper we consider the effects of such an X-ray background on Pop III star formation.  In \RefSec{context} we provide the cosmological context for this study, estimating both the expected intensity of the CXB, and the amount of additional heating required to prevent minihalo collapse. Our numerical methodology is described in \RefSec{methods}, while our results are found in \RefSec{results}.  Finally, our conclusions are gathered in \RefSec{conclusions}.
Throughout this paper we adopt
a $\Lambda$CDM model of hierarchical structure formation, using the following
cosmological parameters: $\Omega_{\Lambda} = 0.7$, $\Omega_{\rm m} = 0.3$, $\Omega_{\rm B} = 0.04$, and $H_0 = 70 \kms \Mpc^{-1}$.

\section{Cosmological Context}
\label{context}
\subsection{Early Cosmic X-ray Background}
\label{HMXB}
The predominant source of X-rays at high redshifts was likely HMXBs.
Supermassive black holes were not yet common during this era, and
while supernovae produce significant X-ray radiation, their transient
nature precludes them from efficiently building up an X-ray
background. We can calculate the energy density \uxr
of X-rays produced by HMXBs as follows:
\begin{equation}
  \uxr(z) = f_{\xrb} \sfrz \left(\frac{\Lxr}{M_{\xrb}}\Delta t_{\xrb}\right) 
  \tH(z) (1+z)^3,
\end{equation}
where $f_{\xrb}$ is the mass fraction of stars that form HMXBs, \sfrz
is the comoving star formation rate density (SFRD) as a function of
redshift, and $\Lxr$, $M_{\xrb}$ and $\Delta t_{\xrb}$ are the
X-ray luminosity, mass and lifespan of a typical HMXB, respectively.  The
X-ray background accumulates over the Hubble time \tH, and the factor
of $(1+z)^3$ accounts for the conversion from a comoving SFRD to a
physical energy density. 

We employ the SFRD calculated by \citet{GreifBromm2006}, but see \citet{Campisietal2011} for a more recent calculation in the context of Pop III gamma-ray burst observations. The \citet{GreifBromm2006} SFRD incorporates both Pop III and Population I/II star formation with self-consistent reionisation and chemical enrichment.  Their SFRD history only extends out to $z=30$; we
extrapolate back to $z=100$ using a simple log-linear fit. 

The mass fraction of stars forming HMXBs, $f_{\xrb}$, is determined by
the mass fraction of stars forming black holes $f_{\textsc{bh}}$, the
fraction of black holes in binary systems $f_{\rm binary}$ and the
fraction of binaries close enough for mass transfer to occur
$f_{\rm close}$. As their IMF is nearly flat with a characteristic
mass of a few $\times10\msun$ \citep{Bromm2013}, we make the
plausible assumption that half of all Pop III stars end up forming
black holes.  Recent work by \citet{StacyBromm2013} found that just
over half of Pop III stars end up in binary pairs; we set $f_{\rm
  binary}$ accordingly. While the orbital distribution of nearby
solar-type stars is well-studied \citep[e.g.,][]{DuquennoyMayor1991},
that of Pop III stars is still very uncertain \citep[but
see][]{StacyBromm2013}.  We therefore assume that the orbital
distribution of Pop III stars is flat, i.e, $dN/d{\rm log} \; a$ is
constant, with a minimum orbital radius of $a=10\rsun$ and a maximum of
$6.5\pc$, which is approximately the size of the self-gravitating baryonic core of a minihalo. For mass transfer onto the BH to occur, the companion star
must at some point exceed its Roche limit; hence we designate all
binaries with orbital distances less than $100\rsun$ as `close.'
Doing so, we find that approximately $2/15$ of Pop III binaries will be
close enough for mass transfer to occur.
 Thus, for Pop III stars, we can estimate $f_{\xrb}$ as follows:  
\begin{equation}
  f_{\xrb} = f_{\textsc{bh}} f_{\rm binary} f_{\rm close} \simeq \left(\frac{1}{2}\right) 
  \left(\frac{1}{2}\right) \left(\frac{2}{15}\right) = \frac{1}{30}.
\end{equation}

Assuming HMXBs accrete material at close to the Eddington limit, we can approximate 
\begin{equation}
  \frac{L_{\xrb}}{M_{\xrb}} \simeq l_\textsc{edd} = 10^{38} \erg \s^{-1} \msun^{-1},
\end{equation}
where $L_{\xrb}$ is the bolometric luminosity.  Following \citet{Jeonetal2014a}, we assume that the spectral distribution of the
emerging radiation takes the form of a thermal multi-colour disc at
frequencies below $0.2 \kev$, and that of a non-thermal power law at
higher frequencies \citep[e.g.,][]{Mitsudaetal1984}.  Assuming the entire accretion luminosity is
emitted between $13.6\ev$ and $10\kev$, the fraction of the total
luminosity emitted between 1 and 10\kev is approximately 30\%.  We
choose 1\kev for the lower limit as the cross section of neutral
hydrogen increases rapidly at lower frequencies.   X-rays below
$\sim1\kev$ thus cannot efficiently contribute to a pervasive
background.  We therefore approximately set
\begin{equation}
\Lxr \simeq \frac{3}{10} L_{\xrb}.
\end{equation}

\begin{figure}
 \begin{center}
   \includegraphics[width=\columnwidth]{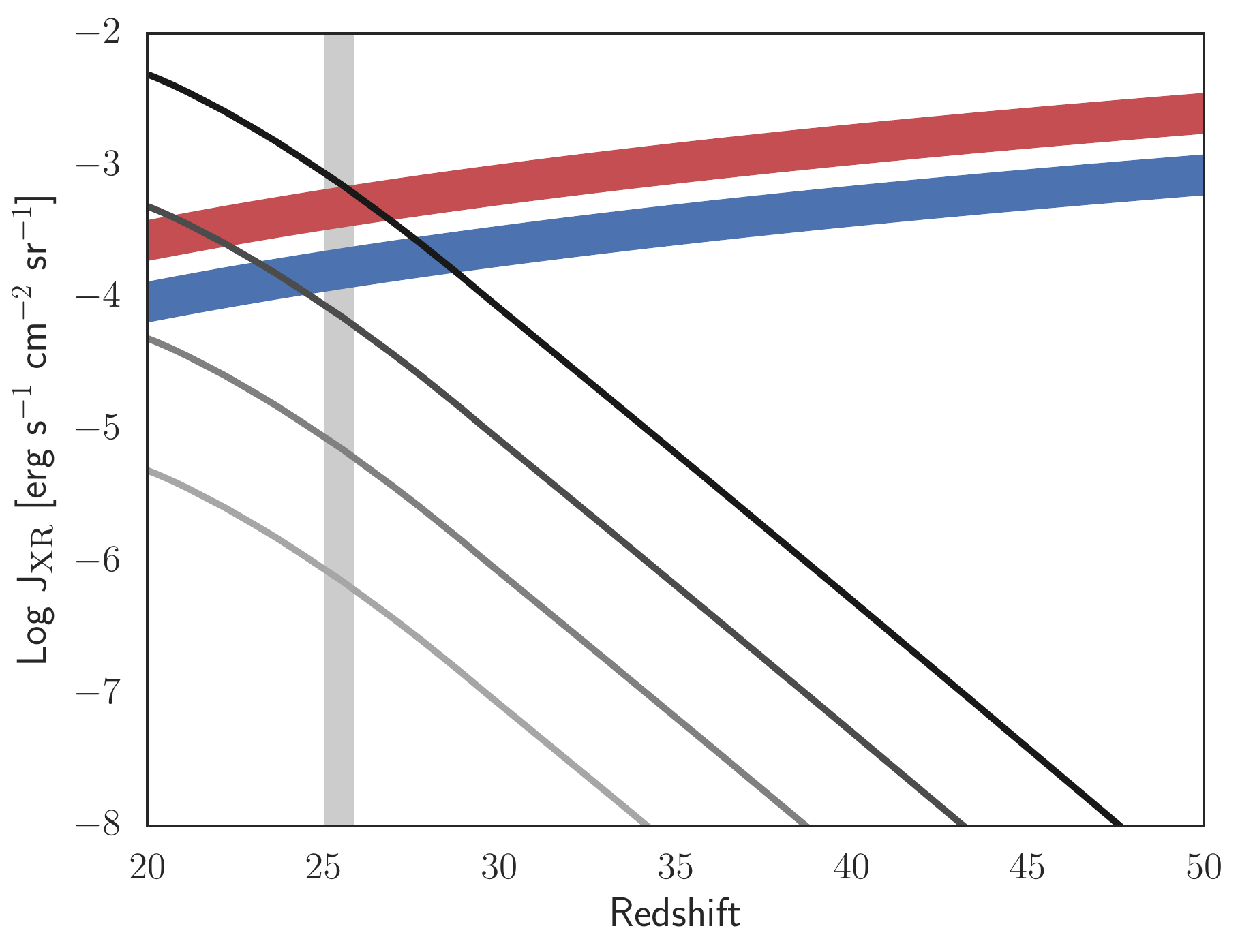}
   \caption{X-ray average intensity as a function of redshift. The
     lightest grey line represents our fiducial model $J_0$, with successively darker grey lines denoting 10, 100, and $1000\,J_0$, respectively. The blue and red ranges denote the critical intensity above which collapse of the baryonic component into a $2\times10^5\msun$ and $10^6\msun$ halo is suppressed, respectively. In each case, the upper limit of the range assumes the contribution from secondary ionisation is negligible, while the lower limit denotes denotes \jcrit for maximally effective secondary ionisation. The vertical grey bar denotes the range of redshifts over which our minihaloes first reach $n=10^{12}\cc$.}
   \label{xrayIntensity}
 \end{center}
\end{figure}

Assuming the typical lifespan of an HMXB is $10^7\yr$ \citep[e.g.,][]{BelczynskiBulikFryer2012, Jeonetal2012}, the average
intensity of this radiation \jxrz is then given by 
\begin{equation} 
  \jxrz =  \int_{\nu_{\rm min}}^{\nu_{\rm max}} \jxrvz d\nu = \frac{c}{4\pi} \uxrz,
\end{equation}
where $h\nu_{\rm min} = 1\kev$ and $h\nu_{\rm max} = 10\kev$.
Following \citet{InayoshiOmukai2011}, we employ
\begin{equation}
  \label{JXRnu}
  \jxrvz = J_{\nu,0}(z) \left(\frac{\nu}{\nu_0}\right)^{-1.5} 
\end{equation}
where $J_{\nu,0}(z)$ is the specific X-ray intensity, $h\nu_0 = 1\kev$, and $J_{\nu, 0}(z)$ is the normalisation factor.
In addition to this fiducial estimate for \jxrz, henceforth referred to as model $J_0$, we consider three
additional models with ten, one hundred and one thousand times the
intensity of $J_0$, as shown in \RefFig{xrayIntensity}. 

\begin{figure*}
\begin{center}
\includegraphics[width=.8\textwidth]{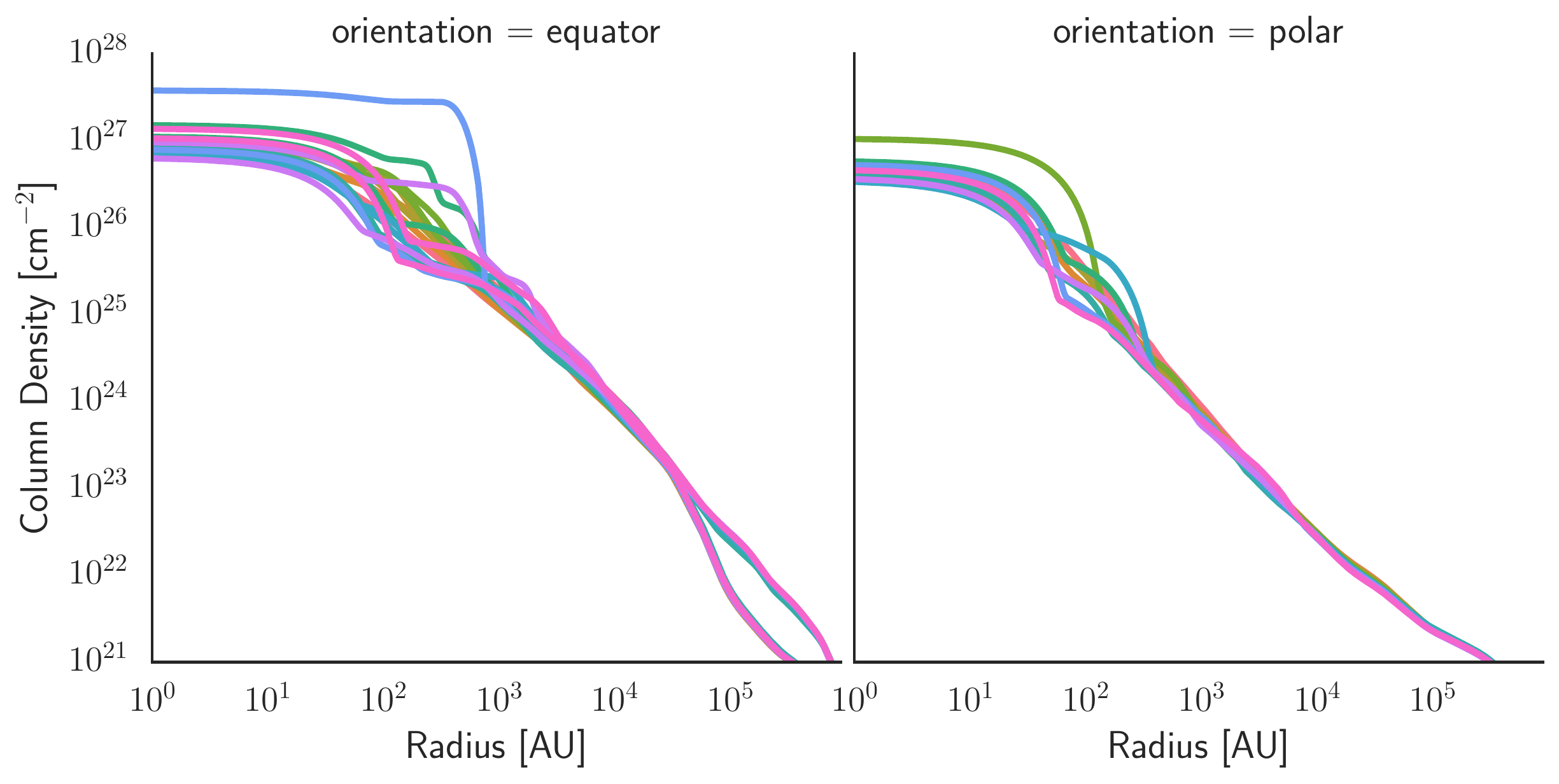}
\caption{\label{fig:column_density} Cumulative column density along both polar and equatorial lines of sight approaching the centre of the minihalo.  Shown is the column density for a selection of snapshots from just prior to the formation of the first sink particle to $5000\yr$ later, with different colors indicating different snapshots.  Note that the scatter in the central $10^3\au$ is larger in the equatorial plane than along the pole: as the accretion disc rotates, occasionally a sink particle orbiting the density-averaged centre lands within the chosen line of sight.}
\end{center}
\end{figure*}

\subsection{Jeans Mass Filtering under X-ray Feedback}
\label{filtering}
The presence of an ionising X-ray background will necessarily heat the
gas in the IGM.  As a result, collapse into a minihalo will be
suppressed when the thermal energy of the gas exceeds the
baryonic gravitational potential energy of the minihalo in question.  The
baryonic gravitational potential energy of a minihalo prior to collapse is given by
\begin{equation}
  \left|\egrav\right| \simeq \frac{\Omega_B}{\Omega_m} \frac{G \mvir^2}{\rvir},
\end{equation}
where $G$ is the gravitational constant, \mvir is the virial mass, and \rvir is the virial radius of the minihalo.
The gas in the minihalo will naturally undergo adiabatic heating as it
is compressed, but the cooling provided by the formation of molecular hydrogen keeps this process from halting collapse.  An additional heating
mechanism, provided here by X-rays, is required to prevent dissipational collapse of the gas.  We approximate this
excess thermal energy \etherm as
\begin{equation}
  \etherm \simeq \gxr \tff \rvir^3,
\end{equation}
where \gxr is the X-ray heating rate per unit volume and \tff is the freefall
time on which gravity draws gas into the minihalo, given by
\begin{equation}
  \tff = \sqrt{\frac{3\pi}{32 G \rho_{\rm b}}},
\end{equation}
where $\rho_{\rm b}$ is the background density. To first order then, the critical heating rate \gcrit for suppressing collapse  is  given by
\begin{equation}
  \gcrit = \frac{\Omega_B}{\Omega_m} \frac{G \mvir^2}{\tff \rvir^4}.
\end{equation}
Utilising the fact that 
\begin{equation}
  \mvir \simeq \frac{4\pi}{3}\rvir^3 (200 \rho_{\rm b})
\end{equation} 
and
\begin{equation}
  \rho_{\rm b} = \rho_0 (1+z)^3,
\end{equation}
where $\rho_0 = 2.5\times10^{-30}\,{\rm g}\cc$ is the average matter density at the present epoch, we can solve for \gcrit as a function of halo mass and redshift:
\begin{equation}
  \begin{split}
    \gcrit(M,z) = \; & 1.3\times10^{-28} \erg\s^{-1}\cc \\  
    &\times \left(\frac{M}{10^6\msun}\right)^{\frac{2}{3}} 
    \left(\frac{1+z}{20}\right)^{\frac{11}{2}},
  \end{split}
\end{equation}
where we have normalised to typical minihalo values.  

Prior to minihalo virialisation, gas densities are relatively low, such that attenuation of the CXB as it penetrates the minihalo is negligible (see \RefFig{fig:khrates}). Given this, \gxr
is related to the intensity of the CXB as follows:
\begin{equation}
\gxr = 4\pi n \int_{\nu_{\rm min}}^{\nu_{\rm max}} \jxrvz \sigma_{\nu}
\left(1 - \frac{\nu_{\rm ion}}{\nu} \right) d\nu + \Gamma_{\textsc{xr}, \rm sec},
\end{equation}
where $n$ is the gas number density and $\Gamma_{\textsc{xr}, \rm sec}$ is the contribution to the heating rate from secondary ionisation. While the contribution of secondary ionisation events to the heating rate has a complicated dependence on the ionisation fraction \citep[][see \RefSec{xrays} for details]{ShullvanSteenberg1985}, $\Gamma_{\textsc{xr}, \rm sec}$ never enhances \gxr by more than a factor of two in the regime considered here.  When secondary ionisation heating is negligible, 
we can similarly estimate the critical X-ray background required to suppress collapse:
\begin{equation}
  \begin{split}
    \jcrit(M,z) = \; & 3.4 \times 10^{-4} \erg\s^{-1} \cm^{-2} \sr^{-1}\\ &\times \left(\frac{M}{10^6\msun}\right)^{\frac{2}{3}}
    \left(\frac{1+z}{20}\right)^{\frac{5}{2}}.
  \end{split}
\end{equation}
When secondary ionisation heating is maximally effective then, \jcrit will be a factor of two lower. Both limits for \jcrit are shown in \RefFig{xrayIntensity} for both a $10^6$  and a $2\times10^5\msun$ minihalo.  These can be interpreted as the approximate range in which the CXB will begin to have a significant impact on the collapse of gas into such a minihalo.

\section{Numerical Methodology}
\label{methods}
\subsection{Initial Setup}
\label{setup}
We use the well-tested $N$-body smoothed particle hydrodynamics (SPH) code \textsc{gadget-2} \citep{Springel2005}. We initialised our simulations in a periodic box of length 140 kpc (comoving) at $z=100$ in accordance with a $\Lambda$CDM model of hierarchical structure formation. An artificially enhanced normalisation of the power spectrum, $\sigma_8 = 1.4$, was used to accelerate structure formation. See \citet{StacyGreifBromm2010} for a discussion of the validity of this choice. High resolution in this simulation was achieved using a standard hierarchical zoom-in technique for both DM and SPH particles. Three nested levels of additional refinement at 40, 30 and 20 kpc (comoving) were added, each centred on the point where the first minihalo forms in the simulation.  As resolution increases, each `parent' particle is replaced by eight `child' particles, such that at the greatest refinement level, each original particle has been replaced by 512 high-resolution particles.  These highest-resolution SPH particles have a mass $m_{\textsc{sph}} = 0.015\msun$, such that the mass resolution of the simulation is $M_{\rm res} \simeq 1.5 N_{\rm neigh} m_{\textsc{sph}} \lesssim 1\msun$, where $N_{\rm neigh} \simeq 32$ is the number of particles used in the SPH smoothing kernel \citep{BateBurkert1997}.

\subsection{Thermodynamics and Chemistry}
\label{chemistry}
Our chemistry and cooling network is the same as that described in \citet{Greifetal2009b}.  We follow the abundances of \h, \hplus, \hminus, \htwo, $\htwo^+$, \he, \heplus, $\he^{++}$, \deut, $\deut^+$, \hd and e$^-$.  All relevant cooling mechanisms, including cooling via \h and \he collisional excitation and ionisation, recombination, bremsstrahlung and inverse Compton scattering are accounted for.  Of particular importance is cooling via the ro-vibrational modes of \htwo, which are excited by collisions with \h and \he atoms and other \htwo molecules.  At high densities, additional \htwo processes must also be included in order to properly model the gas evolution.  For example, three-body \htwo formation and the associated heating becomes important above $n\gtrsim10^8$\cc \citep{Turketal2011}.  The formation rates for these reactions are uncertain; we employ the intermediate rate from \citet{PallaSalpeterStahler1983}. At densities greater than \about$10^9\cc$, the ro-vibrational lines for \htwo become optically thick, decreasing the efficiency of such cooling. We employ the Sobolev approximation and an escape probability formalism to account for this (see \citealt{Yoshidaetal2006, Greifetal2011} for details).

\subsection{Optical Depth Estimation}
\label{attenuation}
\begin{figure}
\begin{center}
\includegraphics[width=1\columnwidth]{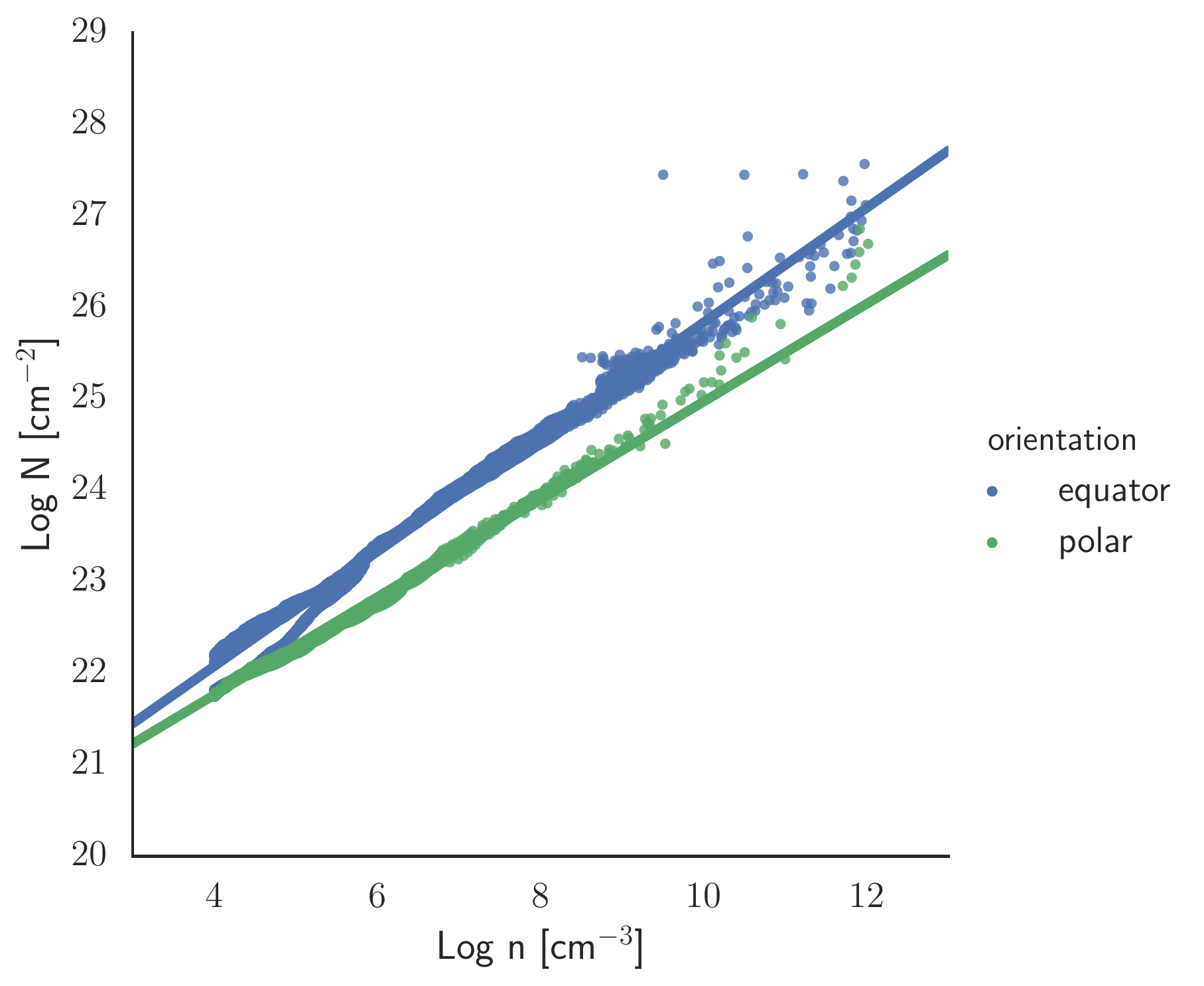}
\caption{\label{fig:ncol_fit}
Column density as a function of gas number density along both polar (green) and equatorial (blue) lines of sight.  Points sample the distribution of column density as a function of number density for their respective lines of sight across several snapshots, while the lines indicate the $\chi^2$ best fit.}
\end{center}
\end{figure}

Over the length scale of our cosmological box, the X-ray optical depth of primordial gas is $\ll 1$ everywhere except approaching the centre of the star-forming minihalo. As the CXB radiation penetrates the minihalo, it will necessarily be attenuated due to the high column density $N$ of the intervening gas.  In order to estimate the optical depth, we directly calculate the column density approaching the centre of the accretion disc along both polar and equatorial lines of sight in the absence of a CXB. As shown in Figure \ref{fig:column_density}, the column density remains essentially constant for the duration of the simulation, and has a simple power-law dependence on radius over several orders of magnitude up to the resolution limit of our simulation ($\sim100\au$). This same power-law behaviour can be seen versus density as well, as shown in Figure \ref{fig:ncol_fit}. In addition we notice that for a given gas density, the column density along the pole is roughly a factor of 10 lower than along the equator.  Performing an ordinary least squares fit to the combined data from several snapshots for $n > 10^4\cc$, we find that the column density along the pole and equator is well fit by 
\begin{equation}
{\rm log}_{10}(N_{\rm \small pole}) = 0.5323\, {\rm log_{10}}(n) + 19.64
\end{equation}
and
\begin{equation}
{\rm log}_{10}(N_{\rm \small eq}) = 0.6262\, {\rm log_{10}}(n) + 19.57, 
\end{equation}
respectively, and does not change appreciably with increasing \jxr.

\begin{figure}
\begin{center}
\includegraphics[width=1\columnwidth]{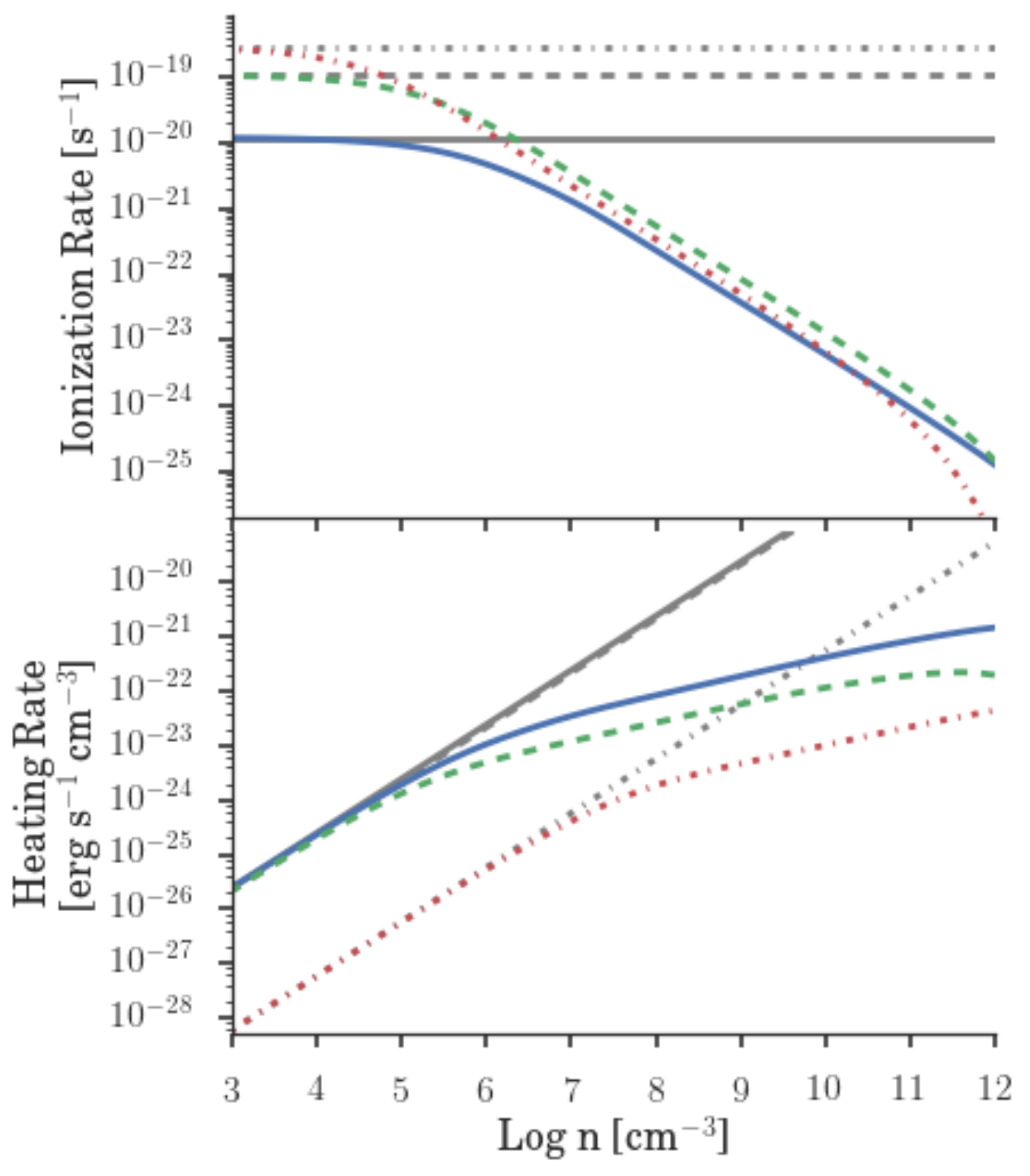}
\caption{\label{fig:khrates}
X-ray ionisation and heating rates for \HI, \HeI and \HeII as a function of total number density for the $\jxr=J_0$ case at $z=25$. In both panels, the solid blue, dashed green, and dash-dotted red lines denote \HI, \HeI, and \HeII, respectively.  The grey lines in each style demonstrate the expected rates in the absence of X-ray self-shielding for that species.  \HI dominates both ionisation and heating when the relative abundances of these species are accounted for, and we see that X-rays penetrating the minihalo experience significant attenuation above densities of $n\sim10^6\cc$.}
\end{center}
\end{figure}

To account for this difference in column density, we assume every line of sight within 45 degrees of the pole \citep{Hosokawaetal2011} experiences column density $N_{\rm \small pole}$ and every other line of sight experiences $N_{\rm \small eq}$. Radiation within the opening angle is attenuated by $e^{-\sigma_{\nu}^i N_{\rm \small pole}}$ while the remainder of the background is attenuated by $e^{-\sigma_{\nu}^i N_{\rm \small eq}}$, allowing us to define an effective optical depth $\tau_{\textsc{xr}}$ such that
\begin{equation}
e^{-\tau_{\textsc{xr}}} = \frac{2 \Omega_{\rm \small pole}}{4\pi} e^{-\sigma_{\nu}^i N_{\rm \small pole}} + \frac{4\pi - 2 \Omega_{\rm \small pole}}{4\pi} e^{-\sigma_{\nu}^i N_{\rm \small eq}},
\end{equation}
\citep[e.g.,][]{ClarkGlover2014}.
Here we use the standard expressions for the photoionisation cross sections $\sigma^i_{\nu}$ of hydrogen and helium \citep[e.g.,][]{BarkanaLoeb2001, OsterbrockFerland2006} and define $\Omega_{\rm \small pole}$ as
\begin{equation}
\Omega_{\rm \small pole} = \int_0^{2\pi}{\rm d}\phi \int_0^{\pi/4}{\rm sin}\theta \,{\rm d}\theta
						 = 1.84\,{\rm sr}.
\end{equation}
Accounting for this attenuation and incorporating the cosmic X-ray background described in Section \ref{HMXB}, the resulting ionisation and heating rates for the $\jxr=J_0$ case at $z=25$ as a function of total number density are shown in Figure \ref{fig:khrates}.

\subsection{X-ray Ionisation and Heating}
\label{xrays}
To study the effects of X-ray ionisation and heating on primordial star formation, we implement a uniform CXB as discussed in \RefSec{HMXB}. For further details, see \citet{Jeonetal2012, Jeonetal2014a}. Accounting for attenuation of the incident X-ray radiation while penetrating the minihalo, the primary ionisation rate coefficient for chemical species $i$ can be written as 
\begin{equation}
k^i_{\rm ion, p} = 4\pi \int_{\nu_{\rm min}}^{\nu_{\rm max}}
\frac{J_{\nu} \sigma^i_{\nu}}{h \nu} e^{-\tau_{\textsc{xr}}} d\nu
\end{equation}
where $\nu_{\rm min} = 1\kev/h$ and $\nu_{\rm max} = 10\kev/h$.  

We include the effects of secondary ionisation from energetic electrons released by the absorption of X-ray photons by adopting the fitting formulae of Shull \& van Steenberg (\citeyear{ShullvanSteenberg1985}; see also \citealt{ValdesFerrara2008, FurlanettoStoever2010}), who calculated the fractions of the initial electron energy going into heating the surrounding gas, as well as into secondary ionisations of \HI and \HeI ($f_{\rm H}$ and $f_{\rm He}$, respectively). While such secondary ionisation events have a significant impact on the ionisation fraction of \HI and \HeI, secondary ionisations of \HeII are negligible \citep{ShullvanSteenberg1985}, and are not included here.  The effective ionisation rates are thus given by
\begin{equation}
k^i_{\rm ion} = k^i_{\rm ion, p} + k^i_{\rm ion, sec}
\end{equation}
where
\begin{equation}
k^{\rm \HI}_{\rm ion, sec} = f_{\rm H} \left( \Gamma_{\rm \HI} + \frac{n_{\rm \HeI}}{ n_{\rm \HI}} \Gamma_{\rm \HeI} \right) \frac{1}{13.6\ev}
\end{equation}
and
\begin{equation}
k^{\rm \HeI}_{\rm ion, sec} = f_{\rm He} \left( \Gamma_{\rm \HeI} + \frac{n_{\rm \HI}}{ n_{\rm \HeI}} \Gamma_{\rm \HI} \right) \frac{1}{24.6\ev}.
\end{equation}
Here $\Gamma_i$ is the heating rate at which excess energy from the initial X-ray photoionisation is released into the gas, given by
\begin{equation}
\Gamma_i = 4\pi n_i \int_{\nu_{\rm min}}^{\nu_{\rm max}} J_{\nu} \sigma^i_{\nu}
\left(1 - \frac{\nu^i_{\rm ion}}{\nu} \right) e^{-\tau_{\textsc{xr}}} d\nu,
\end{equation} 
where $\nu^i_{\rm ion}$ is the ionisation threshold of the species in question; $h\nu_{\rm ion} = 13.6\ev$, 24.6\ev and 54.4\ev for hydrogen, neutral helium and singly ionised helium, respectively.

The fraction of the initial electron energy going into secondary ionisations depends on the hydrogen ionisation fraction $x_{\rm ion, H}$ as follows:
\begin{equation}
f_{\rm H} = 0.3908 \left( 1 - x_{\rm ion, H}^{0.4092} \right) ^{1.7592}
\end{equation}

\begin{equation}
f_{\rm He} = 0.0554 \left( 1 - x_{\rm ion, H}^{0.4614} \right) ^{1.6660}.
\end{equation}
Thus the total heating rate $\Gamma_i^{\rm tot}$, including contributions from both primary and secondary ionizations, can be written as
\begin{equation}
\Gamma_i^{\rm tot} = 4\pi n_i \left(1 - f_i \right)
\int_{\nu_{\rm min}}^{\nu_{\rm max}} J_{\nu} \sigma^i_{\nu}
\left(1 - \frac{\nu^i_{\rm ion}}{\nu} \right) e^{-\tau_{\textsc{xr}}} d\nu.
\end{equation}

\begin{figure*}
  \begin{center}
    \includegraphics[width=0.67\textwidth]{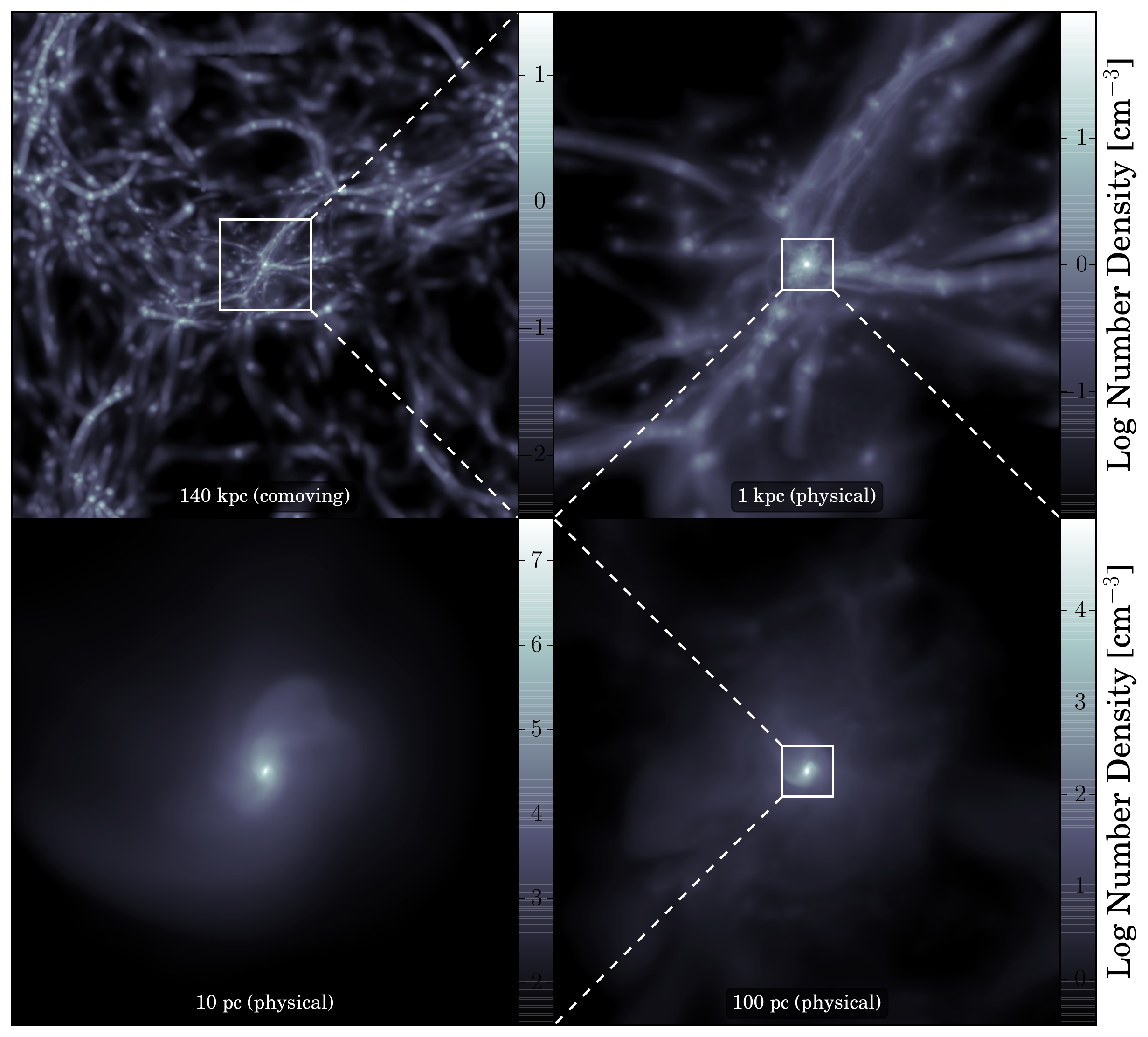}
    \includegraphics[width=0.67\textwidth]{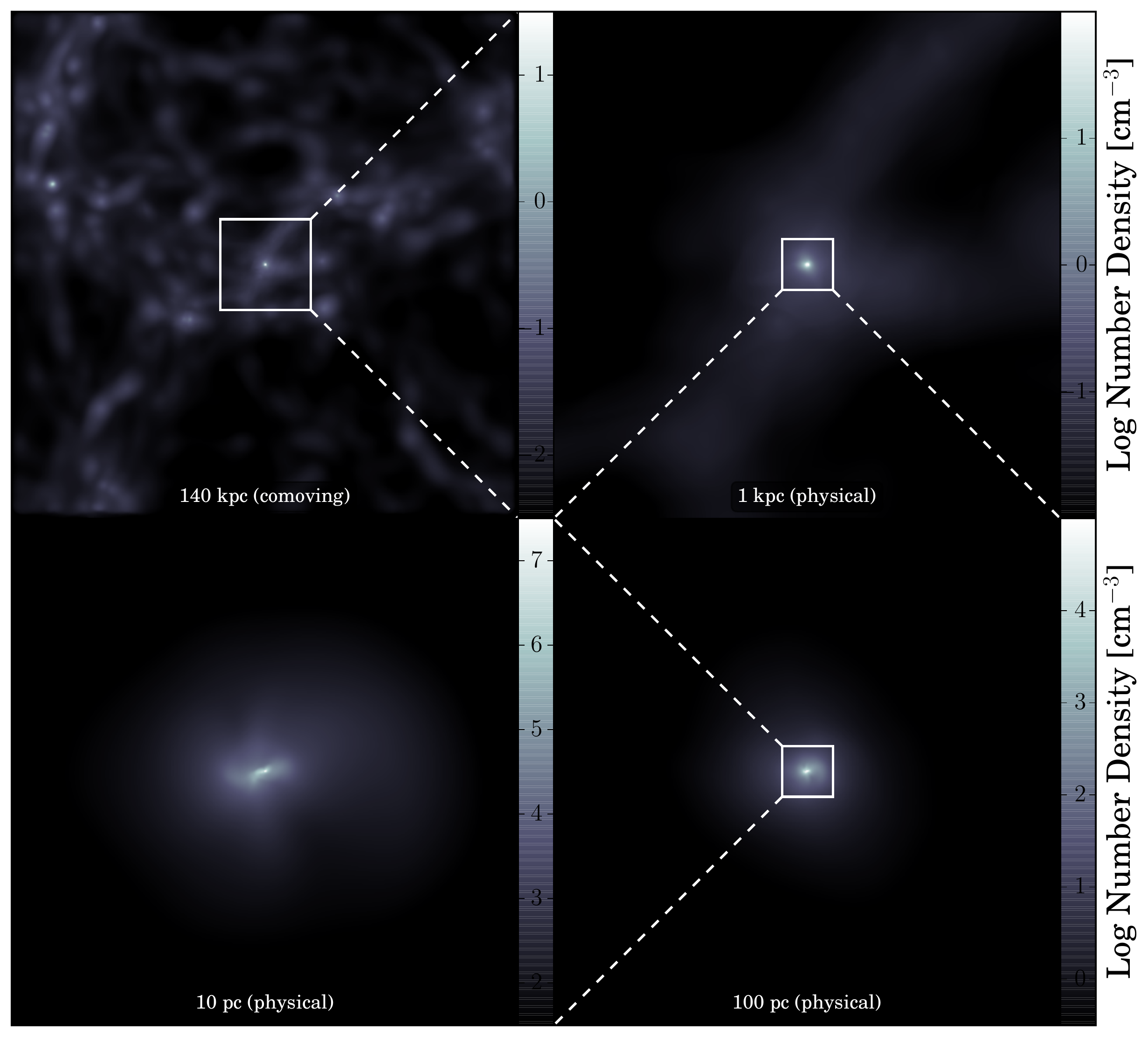}
    \caption{Density projection of the final simulation output 5000\yr after the formation of the first sink particle on progressively smaller scales for both the $\jxr=0$ (top) and the shielded $\jxr=10^2\,J_0$ (bottom) simulations.  White boxes indicate the region depicted on the next smaller scale.  Clockwise from top left: full simulation box; minihalo and surrounding filamentary structure; central 100\pc of minihalo; central 10\pc.  The density scale for each panel is included just to the right -- note that the scaling changes from panel to panel. In both cases, note how the morphology approaches an increasingly smooth, spherical distribution on the smallest scales, where the gas is in quasi-hydrostatic equilibrium.  In the shielded $\jxr=10^2\,J_0$ case, the low-density filamentary structure is smoothed out due to X-ray heating, whereas gas at high densities is shielded and proceeds to collapse unimpeded.}
    \label{zoom-in}
  \end{center}
\end{figure*}

\begin{figure*}
  \begin{center}
    \includegraphics[width=0.9\textwidth]{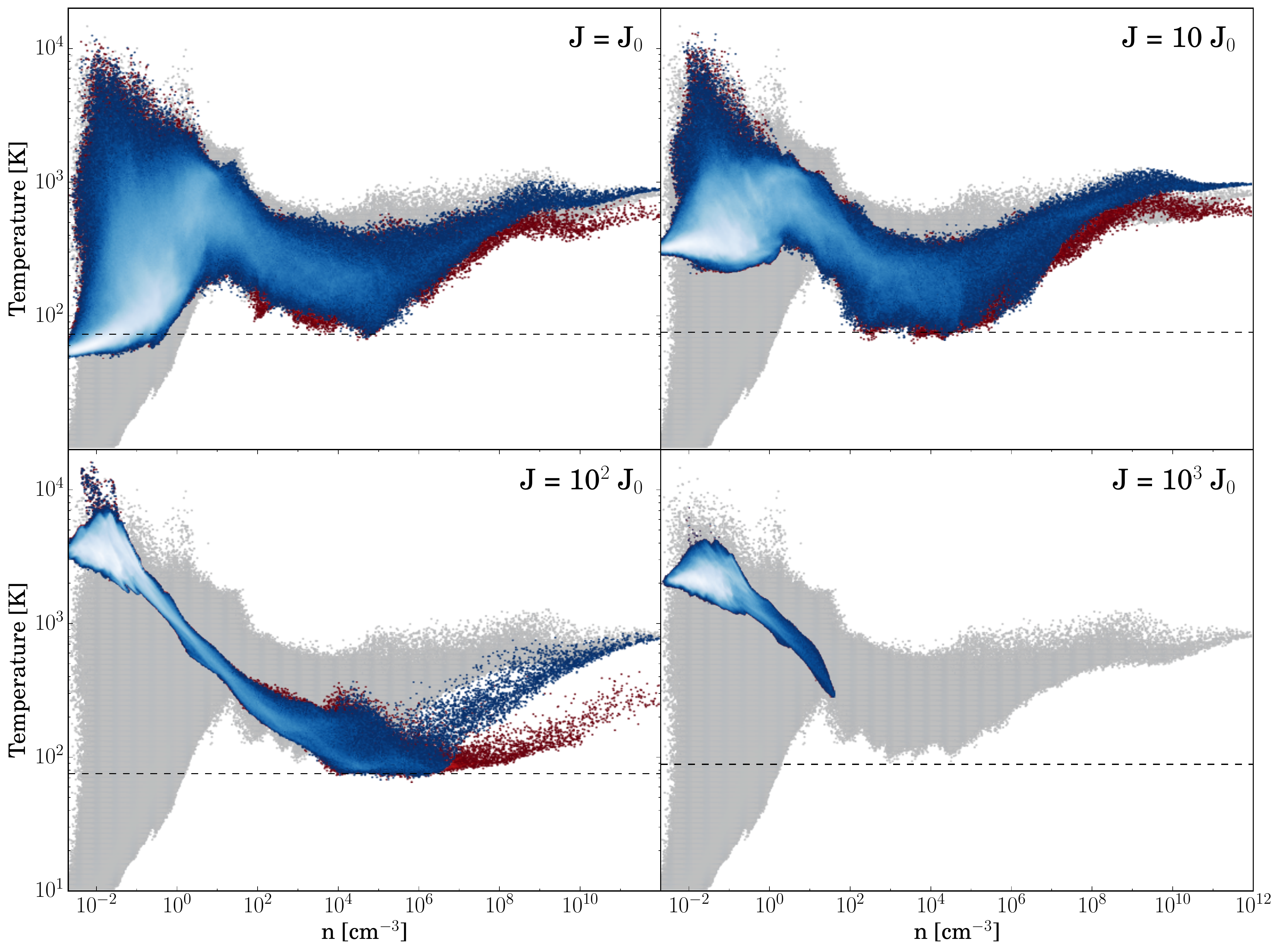}
    \caption{Mass-weighted temperature distribution of the collapsing gas in each simulation, shown just prior to sink formation---or maximum density reached in the case of the $10^3 J_0$ background. Each panel shows the behaviour of the gas in the presence of the CXB, both shielded (in blue) and unshielded (in red). For comparison, the gas behaviour in the $\jxr=0$ case is shown in grey.  Dashed lines denote the CMB temperature. All simulations except the $10^3 J_0$ case successfully collapse to high densities.  Gas at low densities gets progressively hotter and gas in the loitering phase gets progressively cooler with increasing \jxr. Note that in all cases where the minihalo successfully proceeds to collapse, the shielded gas re-converges to the $\jxr=0$ case prior to reaching sink formation densities.}
    \label{Tdens}
  \end{center}
\end{figure*}
\subsection{Sink Particles}
\label{sink_particles}
We employ the sink particle method described in \citet{StacyGreifBromm2010}.  When the density of a gas particle exceeds $n_{\rm max} = 10^{12}\cc$, we replace it and all non-rotationally-supported particles within the accretion radius $r_{\rm acc}$ with a single sink particle.  We set $r_{\rm acc}$ equal to the resolution length of the simulation: $r_{\rm acc} = L_{\rm res} \simeq 50\au$.  Here, 
\begin{equation}
L_{\rm res} \simeq 0.5 \left( \frac{M_{\rm res}}{\rho_{\rm max}} \right)^{1/3},
\end{equation}
where $\rho_{\rm max} = n_{\rm max} m_{\rm H}$.  The sink thus immediately accretes the majority of the particles within its smoothing kernel, such that its mass $M_{\rm sink}$ is initially close to $M_{\rm res} \simeq 1\msun$. Once the sink is formed, additional gas particles and smaller sinks are accreted as they approach within $r_{\rm acc}$ of that sink particle.  After each accretion event, the position and momentum of the sink particle is set to the mass-weighted average of the sink and the accreted particle.

Following the creation of a sink particle, its density, temperature and chemical abundances are no longer updated. The sink's density is held constant at $10^{12}\cc$, and its temperature is kept at 650\kelvin, typical for collapsing gas reaching this density; the pressure of the sink is set correspondingly. Assigning a temperature and pressure to the sink particle in this fashion allows it to behave as an SPH particle. This  avoids the creation of an artificial pressure vacuum, which would inflate the accretion rate onto the sink \citep[see][]{BrommCoppiLarson2002, MartelEvansShapiro2006}. The sink's position and momentum continue to evolve through gravitational and, initially, hydrodynamical interactions with the surrounding particles. As it gains mass and gravity becomes the dominant force, the sink behaves less like an SPH particle and more like a non-gaseous $N$-body particle.

\section{Results}
\label{results}
We perform a total of nine simulations, following each for 5000 years after the formation of the first sink particle in the simulation.  Beyond our fiducial case of $\jxr=J_0$ and the standard case of $\jxr=0$, we examine three additional  simulations with $\jxr = 10$, $10^2$, and $10^3$ times $J_0$. The results of a $\jxr = 10^{-1}\,J_0$ simulation were indistinguishable from $\jxr=0$. Not only is the CXB at high redshifts subject to fairly large uncertainties, but as Pop III star formation is highly biased there is likely a significant amount of cosmic variance.
In addition, we also consider the optically-thin limit where $\tau_{\textsc{xr}} \rightarrow 0$.  While physically unrealistic, these numerical experiments serve as comparison cases to more fully elucidate the physics in the properly shielded simulations.

\subsection{Initial Collapse}
\label{collapse}
\subsubsection{Collapse under Enhanced Cooling}
\label{collapse_acceleration}
\RefFig{zoom-in} shows the final simulation output on various scales for both the $\jxr=0$ and the shielded $\jxr = 10^2\,J_0$ case.  While the expected filamentary structure is visible in all cases, the effects of X-ray heating in the  $\jxr = 10^2\,J_0$ case are readily apparent, with the low-density filamentary gas  
experiencing significant heating. With the usual definition of the virial radius \rvir as $\rho = \rho_{\rm vir} \equiv 178\,\rho_{\rm b}$, where $\rho$, $\rho_{\rm vir}$ and $\rho_{\rm b}$ are the average halo density, the density at the point of virialisation and the background density at the time of halo virialisation, we find that our $\jxr = 0$ minihalo collapses at $z=25.04$ with $\rvir \simeq 85\pc$ and $\mvir \simeq 2.1\times10^5\msun$, typical for the minihalo environment \citep{Bromm2013}.  

\begin{figure}
  \begin{center}
    \includegraphics[width=\columnwidth]{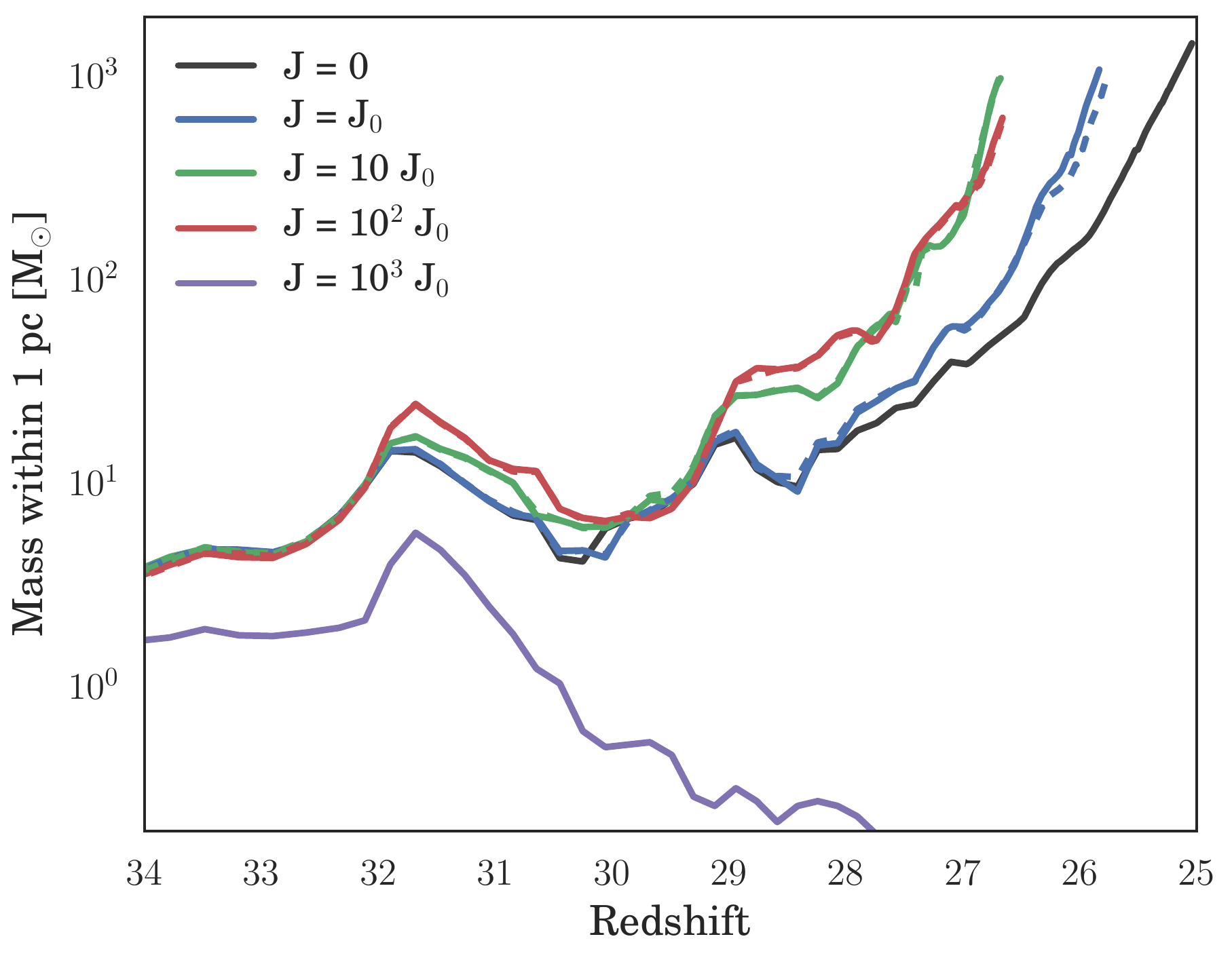}
    \caption{Total gas mass within 1\pc of the highest density point in each simulation over cosmic time. Shielded simulations are denoted by solid lines, unshielded simulations by dashed lines.  Both with and without shielding, as \jxr increases, the gas in the loitering phase becomes cooler. This lowers the Jeans mass, allowing the gas to collapse to high densities sooner.}
    \label{fig:collapse}
  \end{center}
\end{figure}

After the minihalo virialises, the gas continues to collapse in accordance with the standard picture of Pop III star formation \citep[e.g.,][]{StacyGreifBromm2010, Greifetal2012, StacyBromm2013}. This picture is modified slightly in the presence of an X-ray background, but in the $\jxr=0$ case the gas heats adiabatically until reaching $n\sim1\cc$, attaining temperatures of $\sim$1000\kelvin. Beyond this point, the gas is able to cool via the ro-vibrational modes of \htwo, reaching a minimum temperature of \about200\kelvin at a density of $n\sim10^4\cc$.  After exiting the quasi-hydrostatic `loitering phase' \citep{BrommCoppiLarson2002}, the gas enters runaway collapse until $n\sim10^8\cc$, when three-body reactions become important.  This process turns the gas fully molecular by $n\sim10^{12}\cc$, the density at which we form sink particles. As seen in \RefFig{Tdens}, this basic picture holds true even as we vary the strength of the CXB by several orders of magnitude. 

We find that X-ray heating dominates at low densities in all cases, while above $n\sim10^2\cc$ the additional cooling catalysed by X-ray ionisation exceeds it; the precise density at which this transition occurs depends on the strength of the CXB.  This enhanced cooling significantly impacts the subsequent evolution of the gas as it collapses to high densities; the minimum temperature of the gas in the loitering phase approaches the CMB floor as $\jxr$ increases, and---in the unshielded simulations---remains cooler than the $\jxr=0$ case in all later stages of the collapse. This allows the gas to more easily fulfil the Jeans criterion and thus collapse sooner, as demonstrated by \RefFig{fig:collapse}.

When shielding is properly accounted for, the gas becomes optically thick to X-rays as it exits the loitering phase and proceeds to runaway collapse (see \RefFig{fig:khrates}).  In the absence of continued X-ray ionisation, the free electron fraction decays, re-converging with that of the $\jxr=0$ case.  Consequently, the thermodynamic behaviour of the gas at high densities in the shielded simulations is remarkably similar, even as we vary the CXB strength by several orders of magnitude as shown in Figure \ref{Tdens}. This convergence under a wide range of X-ray backgrounds is similar to the behaviour noted by \citet{StacyBrommLoeb2011a} and \citet{Greifetal2011b} in the presence of dark matter--baryon streaming, though we observe an earlier---rather than a later---collapse.

\subsubsection{Collapse Suppression}
 \label{suppression}
\begin{figure}
  \begin{center}
    \includegraphics[width=\columnwidth]{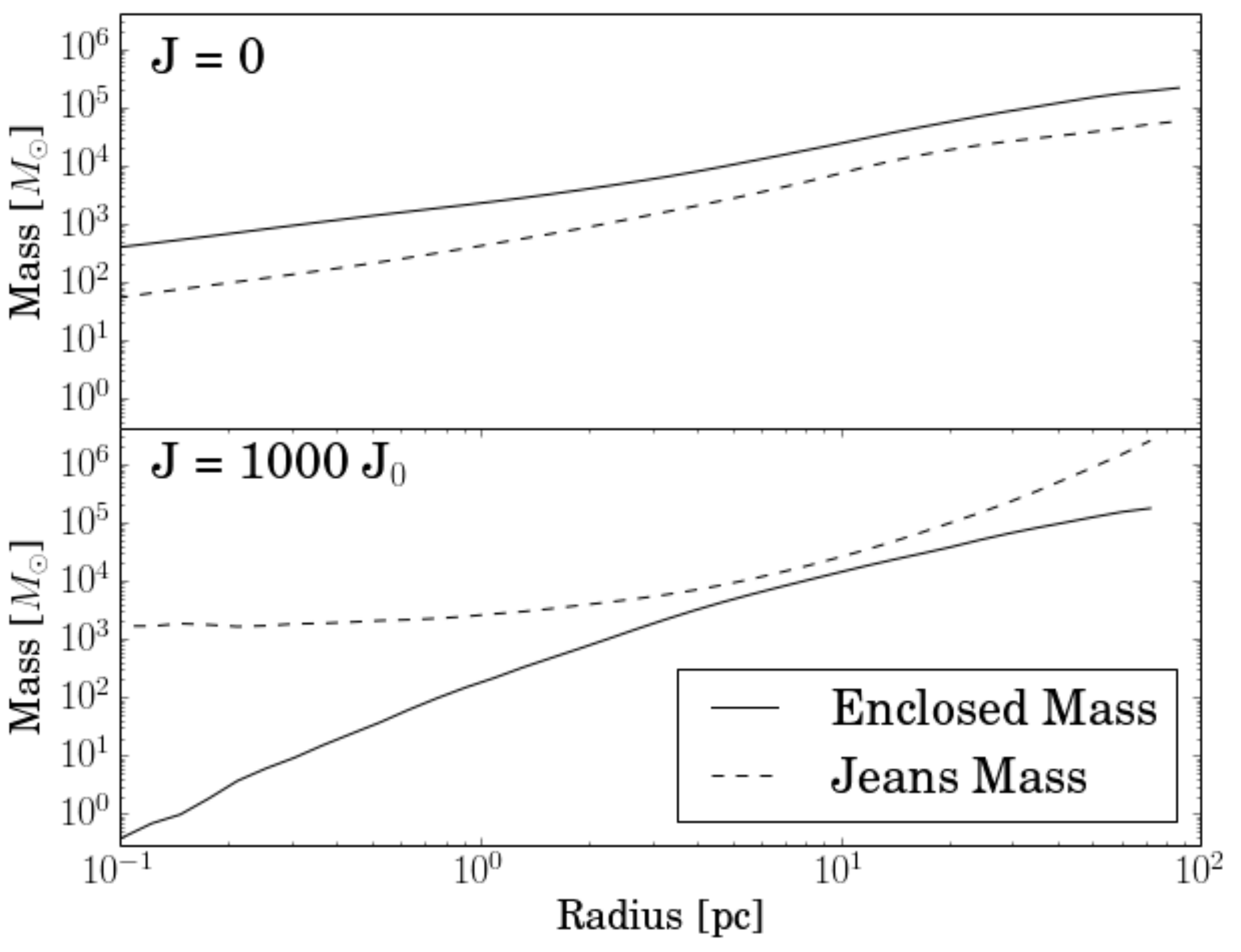}   
    \caption{Shown
      here are the Jeans mass (dotted lines) and total enclosed mass
      (solid lines) within a given radius for  both the $\jxr = 0$ (upper panel) and $\jxr = 1000\,J_0$ case (lower panel). Here we calculate the Jeans mass using the average density and temperature within that radius, extending out to the virial radius of the halo.  The \jxr = 0 case is shown just prior to the formation of the first sink particle; the $\jxr = 1000\,J_0$ case is shown at the maximum density reached over the course of the simulation.}
    \label{suppressed}
  \end{center}
\end{figure}

Minihalo collapse in the $\jxr = 1000\,J_0$ case is completely suppressed.  While the gas in the very centre of the minihalo does initially begin to cool via \htwo, this process is quickly overwhelmed by heating from the increasingly strong X-ray background.  Despite reaching densities and temperatures approaching $n \simeq 50\cc$ and $T \simeq 200\kelvin$ this cold core is eventually dissipated as the CXB continues to heat the gas. 
The reason for this suppression is demonstrated clearly in \RefFig{suppressed}, where we have shown the enclosed mass and the Jeans mass as a function of radius for both the $\jxr = 0$ and the $\jxr = 1000\,J_0$ case. Here, we calculate the Jeans mass using the average density and temperature within a given radius, extending out to the virial radius of the halo.  The $\jxr = 0$ case is shown just prior to the formation of the first sink particle in that simulation; the $\jxr = 1000\,J_0$ case is shown at the maximum density reached over the course of the simulation.  While the enclosed mass in the $\jxr = 0$ minihalo exceeds the Jeans mass on all scales, and thus is universally collapsing, the cold core in the $\jxr = 1000\,J_0$ minihalo never gains sufficient mass to exceed the Jeans criterion.

\subsection{Sink Particle Formation and Accretion}
\subsubsection{Build-up of a Central Disc}
\begin{figure*}
  \begin{center}
    \includegraphics[width=.9\textwidth]{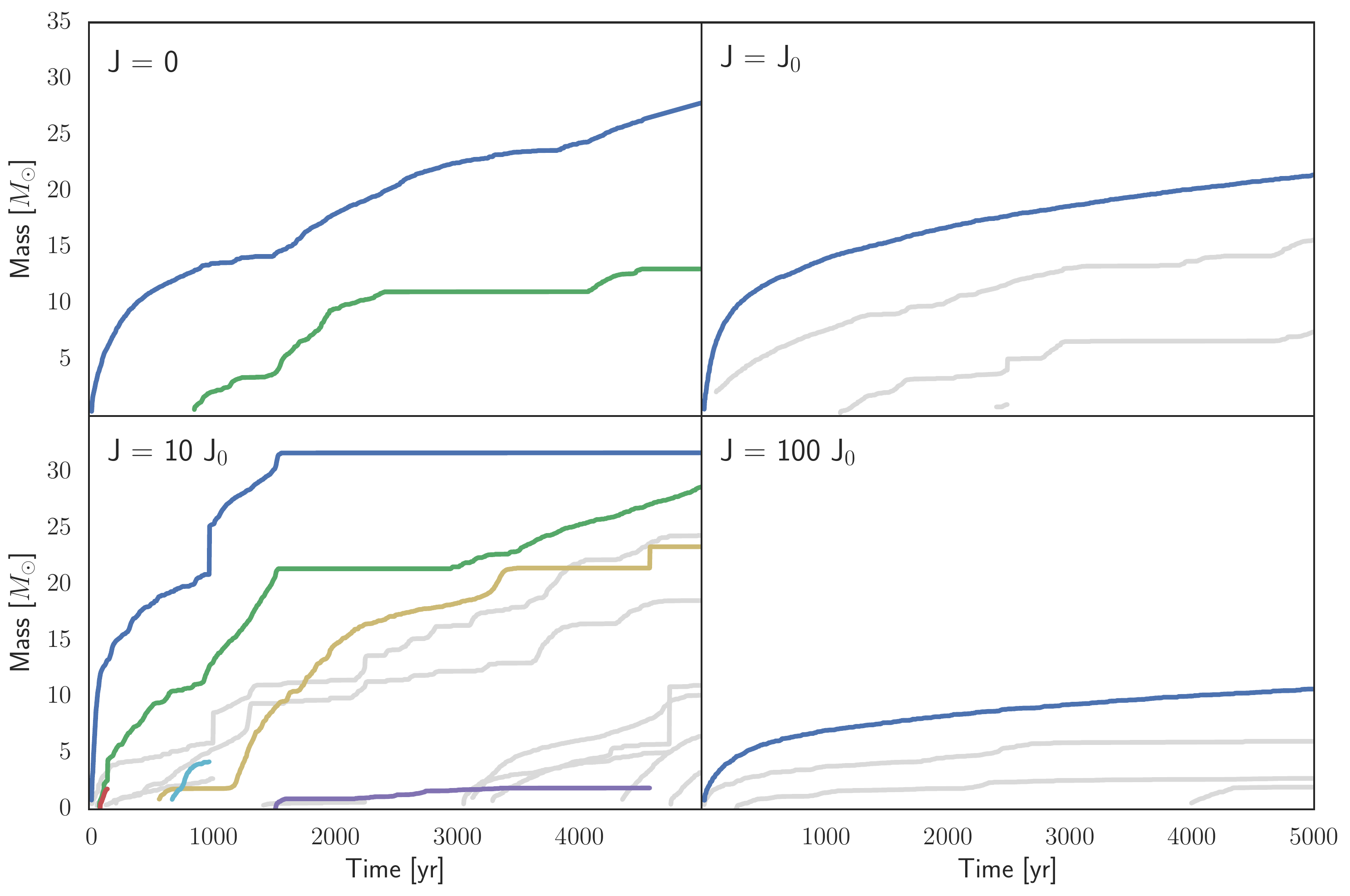}
    \caption{Growth of individual sink particles over time. Sinks for the shielded simulations are coloured throughout in order of formation: blue, green, red, yellow,  cyan, and purple. Lines end where one sink is accreted by another.  The light grey lines show the growth history of sinks in the unshielded simulations for comparison. Note that in the absence of shielding, more sinks form, but the total mass remains roughly constant.}
    \label{sinkmasses}
  \end{center}
\end{figure*}

\begin{figure*}
  \begin{center}
    \includegraphics[width=\textwidth]{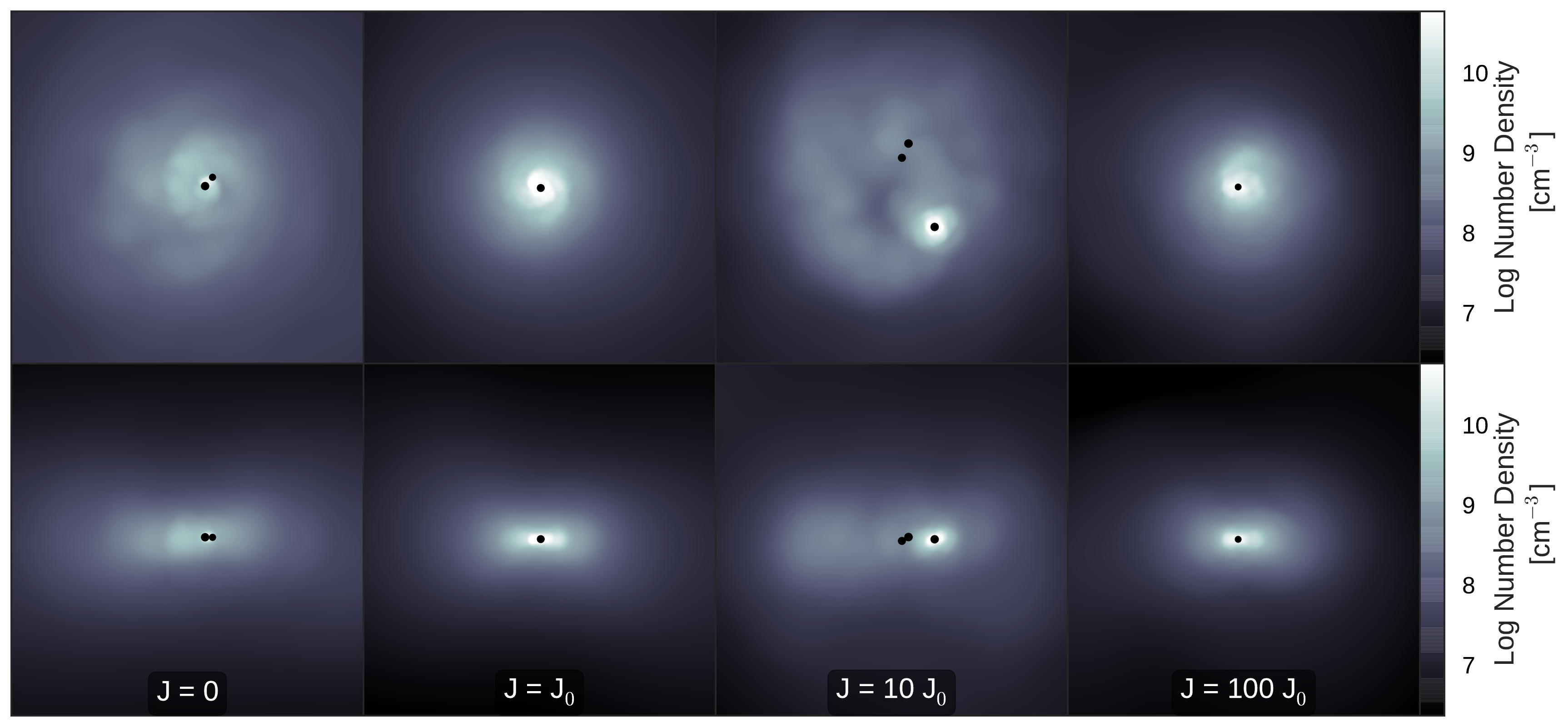}
    \caption{Density projection of the central 10000\au of each simulation 5000\yr after formation of the first sink particle. From left to right: $\jxr=0$, $J_0$, $10\,J_0$, $10^2\,J_0$.  Top row shows the face-on density projection; bottom row shows an edge-on projection.  Black dots mark the location of all sink particles, and scale with the mass of the sink.}
    \label{disks}
  \end{center}
\end{figure*}

\begin{figure}
  \begin{center}
    \includegraphics[width=\columnwidth]{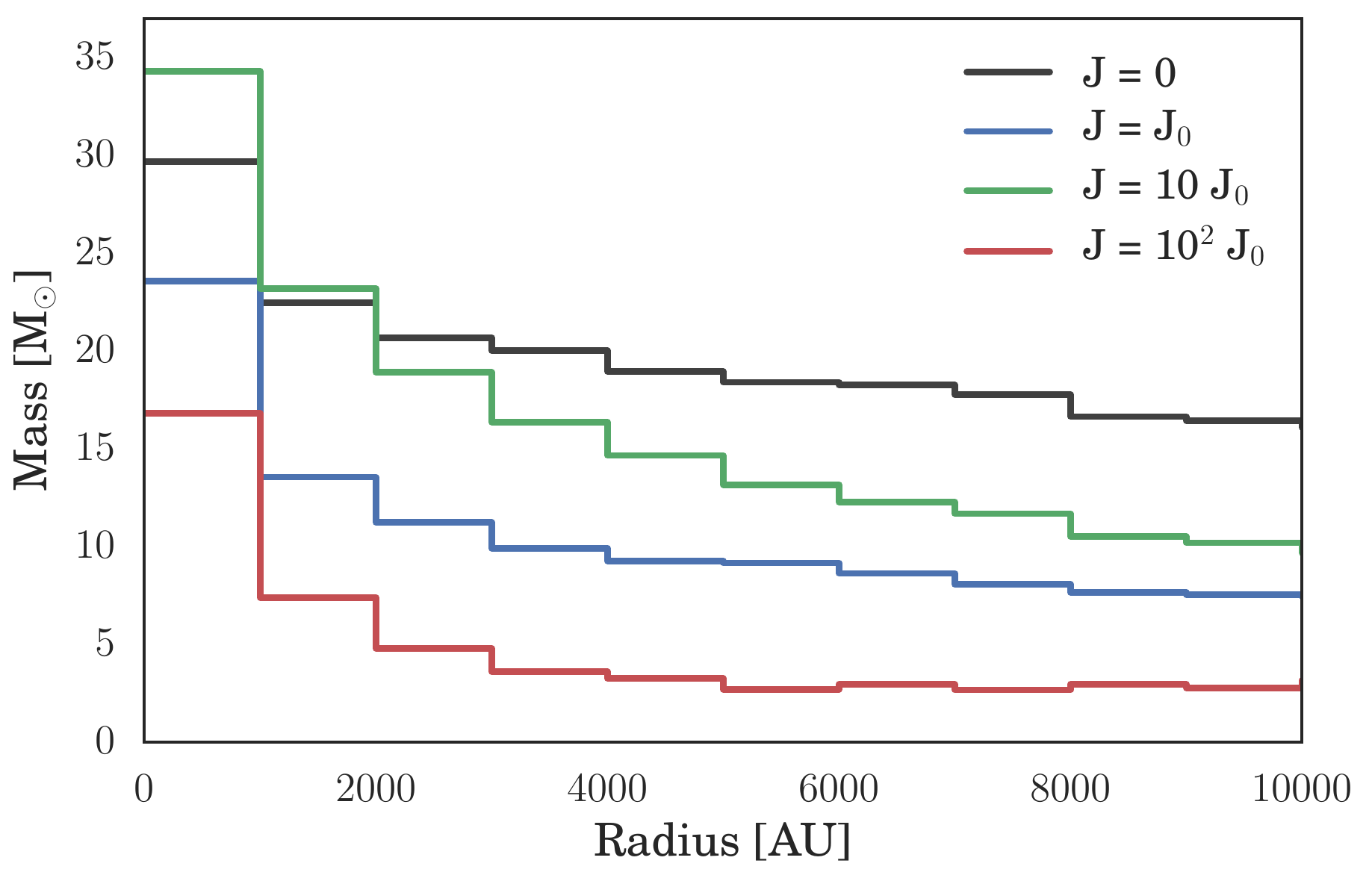}
    \caption{Total gas mass in successive radial bins of 500 AU centred on the highest density point in the simulation.  Shown is the mass distribution just prior to the formation of the first sink particle.}
    \label{mbins}
  \end{center}
\end{figure}

 \RefFig{sinkmasses} shows the growth over time of all sink particles formed in our simulations, from the formation of the first sink to simulation's end 5000\yr later.  Sink particles are formed when the gas in the centre of the minihalo reaches $10^{12}\cc$, and in all cases a central disc forms within $300\yr$ of the first sink particle, as seen in \RefFig{disks}, where we have shown the density structure of the central $10^4\,$AU of each simulation.  In the absence of any X-ray irradiation, the accretion disc fragments and forms a stable binary system, in agreement with previous studies \citep[e.g.,][]{ StacyGreifBromm2010, Clarketal2011a, Clarketal2011b, Greifetal2011, Greifetal2012}. When self-shielding is properly accounted for, the thermodynamic state of the gas in X-ray irradiated minihaloes is remarkably similar to the $\jxr=0$ case by the time it reaches sink formation densities. The mass distribution, however, is in all cases significantly steeper than in the $\jxr=0$ case, as shown in \RefFig{mbins}. This is a direct consequence of the earlier collapse discussed in \RefSec{collapse_acceleration}: the sooner the gas collapses to high densities, the less time is available for gas to accumulate in the centre of the minihalo. 

 Once sink formation densities are achieved, there is no clear trend in the subsequent behaviour of the sink particles formed---either in number or accretion rate.  As seen in \RefTab{masstable}, both the virial mass of the minihalo and the cold core mass (defined here as $n\geq10^2\cc$) decrease with increasing \jxr, as expected given the earlier collapse induced by a stronger CXB.  However, at higher densities the $10\,J_0$ case breaks this trend, suggesting that the final stages of the collapse are somewhat chaotic, and influenced more by small-scale randomness related to turbulence than by the strength of the CXB.

 \begin{table}
  \caption{Total gas mass in various minihalo components for each simulation at $t=5000\yr$. Here we have defined $M_{\rm core}$ and $M_{\rm disc}$ as the total gas mass with $n\geq10^2\cc$ and $n\geq10^8\cc$, respectively. These values are independent of whether shielding is included.}
  \centering
  \begin{tabular}{r c c c c }
    \hline\hline
    \jxr & $M_{\rm vir}$ & $M_{\rm core}$ & $M_{\rm disc}$ & $M_{\rm sink}$\\
    & ($10^4\msun$) & ($10^3\msun$) & (\msun) & (\msun) \\
    \hline
    $0$        & 2.4 & 2.5 & 100 & 41 \\
    $J_0$      & 2.2 & 2.2 &  60 & 23 \\
    $10\,J_0$  & 1.9 & 2.1 & 130 & 74 \\
    $100\,J_0$ & 1.2 & 1.3 &  30 & 11 \\
    \hline
    \\
  \end{tabular}
  \label{masstable}
\end{table}

\begin{figure}
  \begin{center}
    \includegraphics[width=\columnwidth]{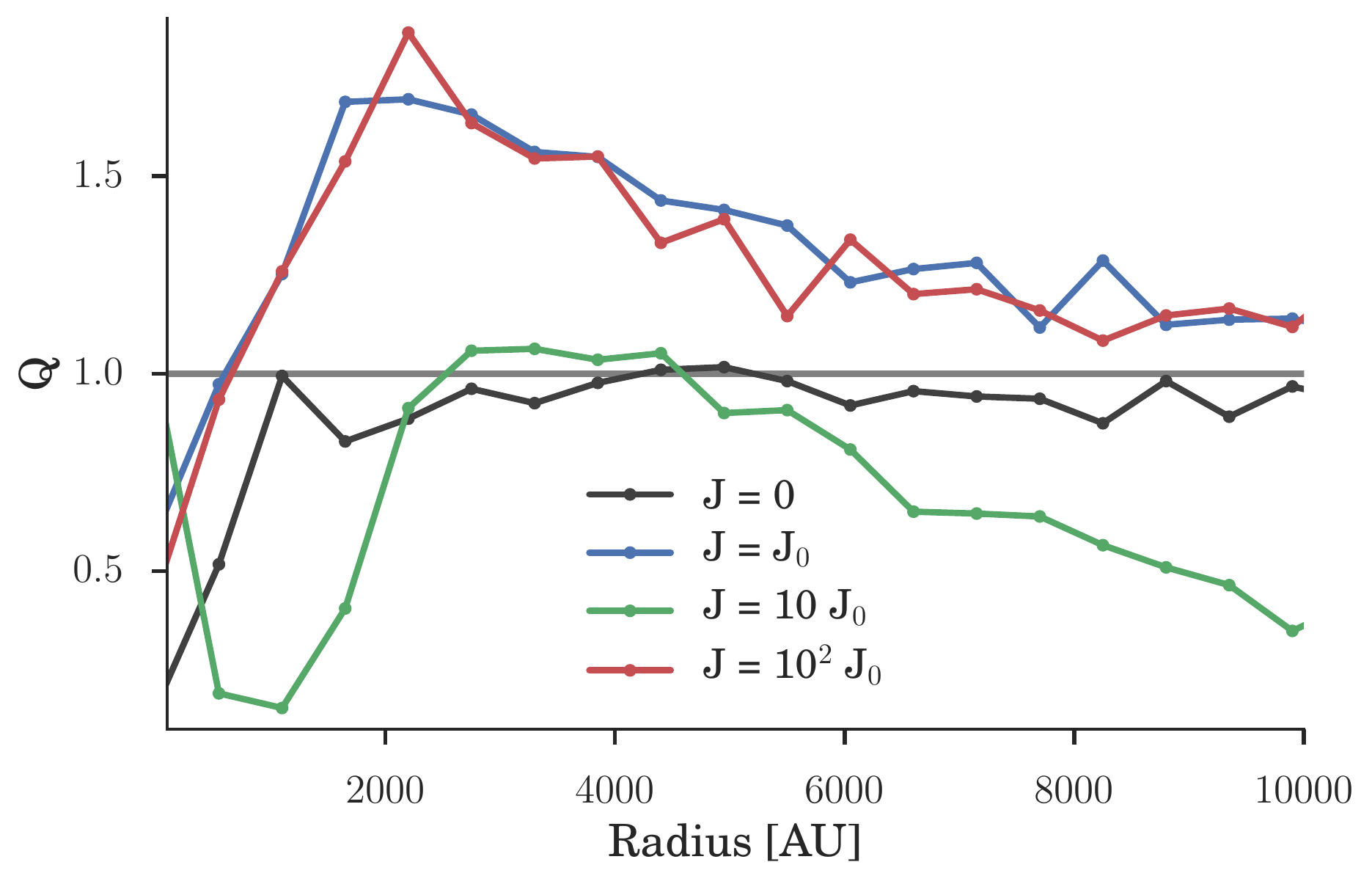}
    \caption{The Toomre Q parameter versus radius shown 5000 yr after the first sink particle forms in each simulation.  The accretion disc is susceptible to fragmentation when $Q\lesssim1$, as is the case for the  $\jxr= 0$ and $10\,J_0$ cases.}
    \label{toomreQ}
  \end{center}
\end{figure}

\subsubsection{Disc Fragmentation}
\label{fragmentation}
In the absence of shielding, the gas invariably fragments, forming a binary or small multiple within $1000\yr$.  When shielding is properly accounted for though, excess cooling of the disc ($n\gtrsim10^8\cc$; see \RefFig{Tdens}) is eliminated, and fragmentation is suppressed.  In fact, only a single sink particle forms in both the $\jxr=J_0$ and $100\,J_0$ cases, and while the $10\,J_0$ simulation still fragments, it does so considerably less than in the absence of shielding.  We may quantify this using the Toomre Q parameter \citep{Toomre1964}: 
\begin{equation}
Q = \frac{c_s \kappa}{\pi G \Sigma}
\end{equation}
where $c_s$ is the gas sound speed, $\kappa$ is the epicyclic frequency of the disc, and $\Sigma$ is the surface density; we replace $\kappa$ with the orbital frequency $\Omega$, as appropriate for Keplerian discs.  While the $Q$ parameter specifically applies to infinitely thin isothermal discs, it is correct to within a factor of order unity when applied to thick discs \citep{Wangetal2010}, as is the case here. \RefFig{toomreQ} shows the $Q$ parameter evaluated in mass-weighted spherical shells centred on the accretion disc $5000\yr$ after the first sink particle has formed.  As the mass within these shells is dominated by the disc component, applying this analysis to the disc particles alone would have a negligible impact on the results \citep[e.g.,][]{Greifetal2012}.  The $\jxr=0$ and $10\,J_0$ simulations maintain $Q\lesssim1$ throughout the disc, and are thus susceptible to fragmentation.  On the other hand, save for the central few hundred AU, where they approach the resolution limit of the simulation, the $J_0$ and $100\,J_0$ discs stay well above $Q=1$, hence the lack of fragmentation.

\begin{figure}
  \begin{center}
    \includegraphics[width=\columnwidth]{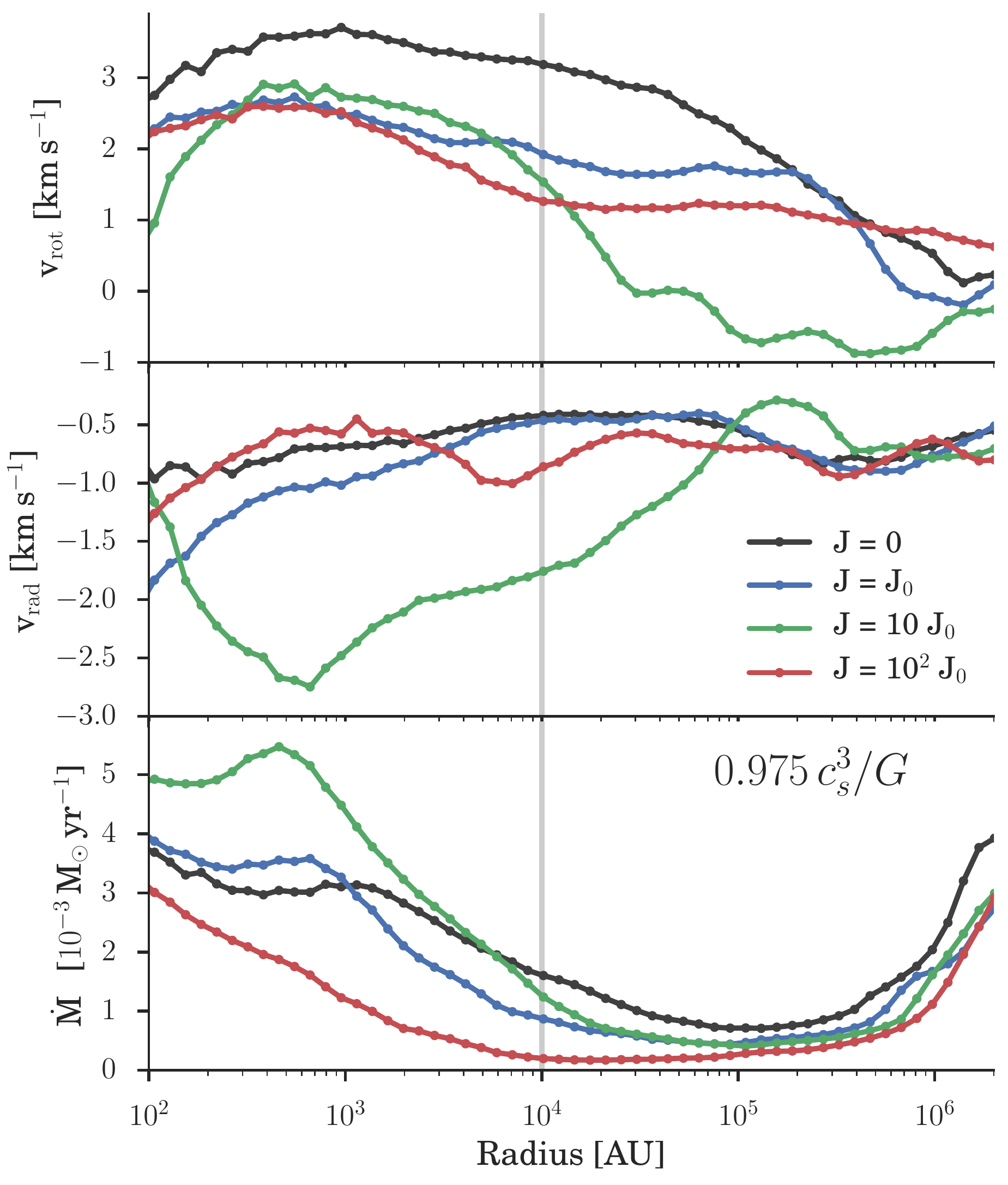}
    \caption{From top to bottom: the rotational velocity, radial velocity, and Shu accretion rate versus distance to the centre of the minihalo for each simulation as denoted in the legend.  The vertical gray line at $10^4\au$ marks the approximate limit of the accretion disc, and the limit to which Figures \ref{disks}, \ref{mbins} and \ref{toomreQ} are displayed.  Note the significant discrepancies in the behaviour of the $\jxr=10\,J_0$ case in each panel.}
    \label{core_profile}
  \end{center}
\end{figure}
While the $\jxr=10\,J_0$ minihalo also collapses early, in the same manner as the $J_0$ and $100\,J_0$ simulations, it still experiences significant fragmentation.  This is primarily due to the specific details of its collapse history, modulated by small-scale turbulence, rather than the precise value of \jxr, and can be understood by examining the velocity profile of the minihalo, shown in \RefFig{core_profile} along with the Shu accretion rate \citep{Shu1977} for mass-weighted spherical bins out to $\sim$$10\pc$. While the inner regions of the $10\,J_0$ accretion disc exhibit approximately Keplerian rotation in line with the other simulations, the gas at large radii experiences significantly less rotational support.  This results in a much larger infall velocity, leading in turn to a higher accretion rate.  This overwhelms the ability of the accretion disc to support itself against the growth of perturbations, resulting in significant fragmentation.

\section{Summary and Conclusions}
\label{conclusions}
We have performed a suite of cosmological simulations employing a range of CXB models, focusing on the impact of such a background on Pop III stars forming in a minihalo.  As three-body processes turn the gas fully molecular by $n=10^{12}\cc$, following the evolution of the gas up to this density allows us to fully capture the impact of a CXB on \htwo and \hd cooling in the gas.  After the gas reaches $10^{12}\cc$ we form sink particles, enabling us to study the subsequent evolution of the system, which we follow for an additional 5000\yr.

X-rays have two competing effects on primordial gas, as ionising the gas serves to both heat it and increase the number of free electrons, which catalyse the formation of molecular hydrogen.  As \htwo is the main coolant in primordial gas, this actually enhances cooling.  We find that heating dominates in low density gas, but is overwhelmed at higher densities by the enhanced cooling X-rays provide.  The transition between these two regimes occurs between $n=1$ and 100\cc, depending on the strength of the CXB.  

Previous work investigating the impact of a CXB on structure formation in the early universe has generally been found to increase the supply of cold gas in a given halo \citep{HaimanAbelRees2000, VenkatesanGirouxShull2001, GloverBrand2003, Cen2003, KuhlenMadau2005, Jeonetal2012}.  In particular, \citet{KuhlenMadau2005} found the increase in cold gas available for star formation was greatest in $2\times10^5$--$10^6\msun$ minihaloes, where the increase could exceed 1--2 orders of magnitude for moderate X-ray ionisation rates. This comports with our findings---allowing the X-ray irradiated haloes to evolve to the same cosmic time as the $\jxr=0$ minihalo would result in a relatively larger supply of cold dense gas. As X-rays penetrating the minihalo begin to experience significant attenuation above $n\sim10^6\cc$ though, the primary impact of a CXB is on gas in the loitering phase and below. While X-ray heating dominates below $n\sim 1\cc$, minihaloes that overcome this impediment collapse earlier, as the cooler gas in the loitering phase requires a smaller Jeans mass to proceed to high densities.  

The X-ray background is severely attenuated by the time it reaches sink formation densities. As a result its impact on the gas temperature is largely neutral, with the subsequent behaviour of the sink particles formed influenced more by small-scale  turbulence than the strength of the CXB.  Consequently, the characteristic mass of the stars formed is quite stable even as we vary the CXB strength by several orders of magnitude, and does not change dramatically even when the supply of cold gas in the centre of the minihalo is significantly increased, as in the $10\,J_0$ case.  Instead, this causes the disc to fragment, forming several protostellar cores.

It should be noted that these findings are somewhat sensitive to the density at which the minihalo becomes opaque to X-rays.  While we found no difference in the total column density between simulations, a more robust estimate of the local column density would be beneficial \citep[e.g.,][]{Hartwigetal2015, Safranek-Shraderetal2012}. Additionally, there is a possibility that the more rapid minihalo collapse induced by a strong CXB may have an impact on the velocity dispersion of the infalling gas, suppressing fragmentation and possibly increasing the mass of the stars formed. While our simulations lack sufficient resolution to verify this, the findings of \citet{Clarketal2011a} suggest that stars forming in pre-ionised minihaloes experience more turbulence, resulting in a somewhat larger characteristic mass than in pristine minihaloes. Finally, our findings may have implications for reionisation and the 21-cm signal \citep{FurlanettoPengBriggs2006, Mirocha2014}: while the impact on the number of fragments and characteristic mass of Pop III appears to be nearly neutral, low-density gas is still smoothed by X-ray heating, thus resulting in a lower IGM clumping factor.

\section*{Acknowledgements}
The authors thank Paul Clark for many insightful comments, and acknowledge the Texas Advanced Computing Center (TACC) at The University of Texas at Austin for providing HPC resources under XSEDE allocation TG-AST120024. This study was supported in part by NSF grants AST-1009928 and AST-1413501, and by the NASA grant NNX09AJ33G. A.S. gratefully acknowledges support through NSF grants AST-0908553 and AST-1211729. J.H. thanks Chalence Safranek-Shrader for many enlightening conversations. This research has made use of NASA's Astrophysics Data System and Astropy, a community-developed core Python package for Astronomy \citep{Robitailleetal2013}.


\bibliography{../../biblio}
\bibliographystyle{mn2e_fixed}  

\end{document}